\documentclass[prd,aps,superscriptaddress,10pt,nobibnotes,notitlepage,twocolumn,longbibliography,nofootinbib,preprintnumbers,dvipsnames]{revtex4-2}

\usepackage{cool}
\usepackage[normalem]{ulem}
\usepackage{sidecap}
\usepackage[latin1]{inputenc}
\usepackage{graphicx}
\usepackage{float}
\usepackage{latexsym}
\usepackage{graphicx}
\usepackage{amssymb}
\usepackage{amsmath,mathtools}
\usepackage{amsfonts}
\usepackage{tikz} 
\usepackage{ragged2e}
\usepackage{stackengine}
\usepackage{scalerel}

\usepackage[colorlinks=true,citecolor=blue,hyperfootnotes=false]{hyperref}

\usepackage{dcolumn}
\usepackage{textcomp}
\usepackage{xfrac}
\usepackage{slashed}
\usepackage{multirow}

\def\be{\begin{equation}}
\def\ee{\end{equation}}
\def\figs/B{B}
\def\bea{\begin{eqnarray}}
\def\eea{\end{eqnarray}}
\def\bg{\begin{eqnarray}}
\def\nd{\end{eqnarray}}

\def\ln{{\rm log}}

\def\beq{\begin{equation}}
\def\eeq{\end{equation}}

\usepackage{graphicx}
\usepackage{dcolumn}
\usepackage{bm}
\usepackage{hyperref}
\usetikzlibrary{decorations.pathmorphing}
\tikzset{snake it/.style={decorate, decoration=snake}}

\begin{document}

\preprint{MIT-CTP/5581}

\title{Higgs Criticality beyond the Standard Model}

\author{Thomas Steingasser}
 \email{tstngssr@mit.edu}
\affiliation{%
Department of Physics, Massachusetts Institute of Technology, Cambridge, MA 02139, USA }
\affiliation{Black Hole Initiative at Harvard University, 20 Garden Street, Cambridge, MA 02138, USA
}%

\author{David I. Kaiser}
 \email{dikaiser@mit.edu}
\affiliation{%
Department of Physics, Massachusetts Institute of Technology, Cambridge, MA 02139, USA
}%

\date{\today}

\begin{abstract}
Both parameters in the Higgs field's potential, its mass and quartic coupling, appear fine-tuned to near-critical values, which gives rise to the hierarchy problem and the metastability of the electroweak vacuum. Whereas such behavior appears puzzling in the context of particle physics, it is a common feature of dynamical systems, which has led to the suggestion that the parameters of the Higgs potential could be set through some dynamical process. In this article, we discuss how this notion could be extended to physics beyond the Standard Model (SM). We first review in which sense the SM Higgs parameters can be understood as near-critical and show that this notion can be extrapolated in a unique way for a generic class of SM extensions. Our main result is a prediction for the parameters of such models in terms of their corresponding Standard Model effective field theory Wilson coefficients and corresponding matching scale. For generic models, our result suggests that the scale of new (bosonic) physics lies close to the instability scale. We explore the potentially observable consequences of this connection, and illustrate aspects of our analysis with a concrete example. 
Lastly, we discuss implications of our results for several mechanisms of dynamical vacuum selection associated with various Beyond-Standard-Model constructions.
\end{abstract}

\maketitle


\section{Introduction}

Modern particle physics can be understood in terms of the principles of \textit{symmetry}, \textit{unitarity}, and \textit{naturalness}. The former two have been essential in the construction of the Standard Model (SM) in general and the prediction of the Higgs boson in particular, while the latter has played a crucial role in the search for theories beyond this framework. Yet the data collected since the discovery of the Higgs boson seem to indicate that these celebrated principles might fail to explain the properties of the Higgs field itself.
Most important, the idea of naturalness seems to be in conflict with the surprising degree of fine-tuning of both parameters in the Higgs field's potential: the mass term and the quartic self-coupling~\cite{Buttazzo:2013uya,Giudice:2017pzm,Giudice:2021zgc,Buttazzo:2013rxi,Espinosa:2018mfn,Steingasser:2022yqx,Espinosa:2007zz,Espinosa:2007zza}.\footnote{This simple observation could be extended to the cosmological constant problem, which could be understood as the constant part of the Higgs potential being fine-tuned against the purely gravitational term. See also Ref.~\cite{Lee:2023wdm}.} 

Perhaps as remarkable as the apparent tuning of these parameters is the observation that they seem to be fine-tuned to \textit{near-critical values}, that is, they lie close to special values marking the transition between qualitatively different types of potentials. Due to the smallness of the mass parameter, it lies close to the transition from a strictly convex potential near its origin to one with a non-trivial minimum. Meanwhile, the quartic coupling's running towards small (negative) values at high energies can also be understood as the trajectory of its running under renormalization-group (RG) flow being close to the transition point from a stable, convex potential to one that is unstable and unbounded from below. Although highly unusual in the context of particle physics, such behavior is a common feature of dynamical systems~\cite{PhysRevLett.59.381}. 

It thus appears plausible that some physical mechanism has tuned the Higgs field's mass and quartic coupling towards critical values. Several authors have explored dynamical scenarios that could account for this near-critical behavior~\cite{Giudice:2021viw,Khoury:2019yoo,Feldstein:2006ce,Giudice:2019iwl,Dvali:2004tma,Dvali:2003br,Eroncel:2018dkg,Arkani-Hamed:2020yna,Csaki:2020zqz,Strumia:2020bdy,Cheung:2018xnu,Graham:2015cka,Nelson:2017cfv,Gupta:2018wif,TitoDAgnolo:2021pjo,TitoDAgnolo:2021nhd,Trifinopoulos:2022tfx,Cline:2018ebc,Khoury:2021zao,Kartvelishvili:2020thd,Khoury:2019ajl}. If some mechanism has tuned the parameters of the Higgs potential in the SM to near-critical values, this raises the question whether and how such mechanisms might relate to yet undiscovered Beyond-Standard-Model (BSM) physics.  

Of course, it is impractical to investigate potential tuning mechanisms across the huge range of proposed BSM models. Instead we consider the possibility that whatever mechanism causes the mass and quartic coupling parameters within the SM Higgs potential to be tuned to near-critical values {\it also} yields near-critical behavior of the full theory underlying the SM itself. In other words, we explore scenarios in which some dynamical mechanism yields a coordination among apparently unrelated parameters, such that the Higgs potential lies near a critical configuration separating distinct phases within the full theory.

Such scenarios may be investigated in a remarkably efficient way and with reasonable accuracy by employing standard techniques from effective field theory (EFT). In this work, we consider the possibility that the Higgs sector is extended at some large energy scale $\Lambda$. At low energies, the effects of such new physics can then be incorporated into the SM through the addition of higher-dimensional operators, such as corrections to the potential of the form
\begin{align}\label{I/Vexta}
    V_{\rm eff}(\mu,H)= & - \frac{1}{4} m_{\rm eff}^2(\mu,H)  H^2  + \frac{1}{4} \lambda_{\rm eff}(\mu,H) H^4 + \nonumber \\ 
    &+ \frac{C_6}{\Lambda^2} H^6 + ... ,
\end{align}
where $\mu$ is the scale at which the effective potential is evaluated.\footnote{Throughout this article, we consider the behavior of the zero-temperature effective potential. The quantity $H$ is defined through $H^2= 2 \mathbf{H}^2$, where $\mathbf{H}$ is the full Higgs doublet, including the wave-function renormalization factor.} 

As long as we are interested in field values smaller than the cutoff scale $\Lambda$, the properties of the SM extension of interest can be entirely understood to leading order through the combination $C_6/\Lambda^2$, including the effects of this new term on the mass parameter and the quartic coupling. Having established that the latter two appear fine-tuned to lie close to critical values corresponding to phase transitions, Eq.~\eqref{I/Vexta} suggests that the relevant question becomes:
\textit{How would mechanisms driving the Higgs potential's parameters towards critical values affect the coefficient $C_6/\Lambda^2$?} 

\begin{figure}[t!]
    \centering
    \includegraphics[width=0.45\textwidth]{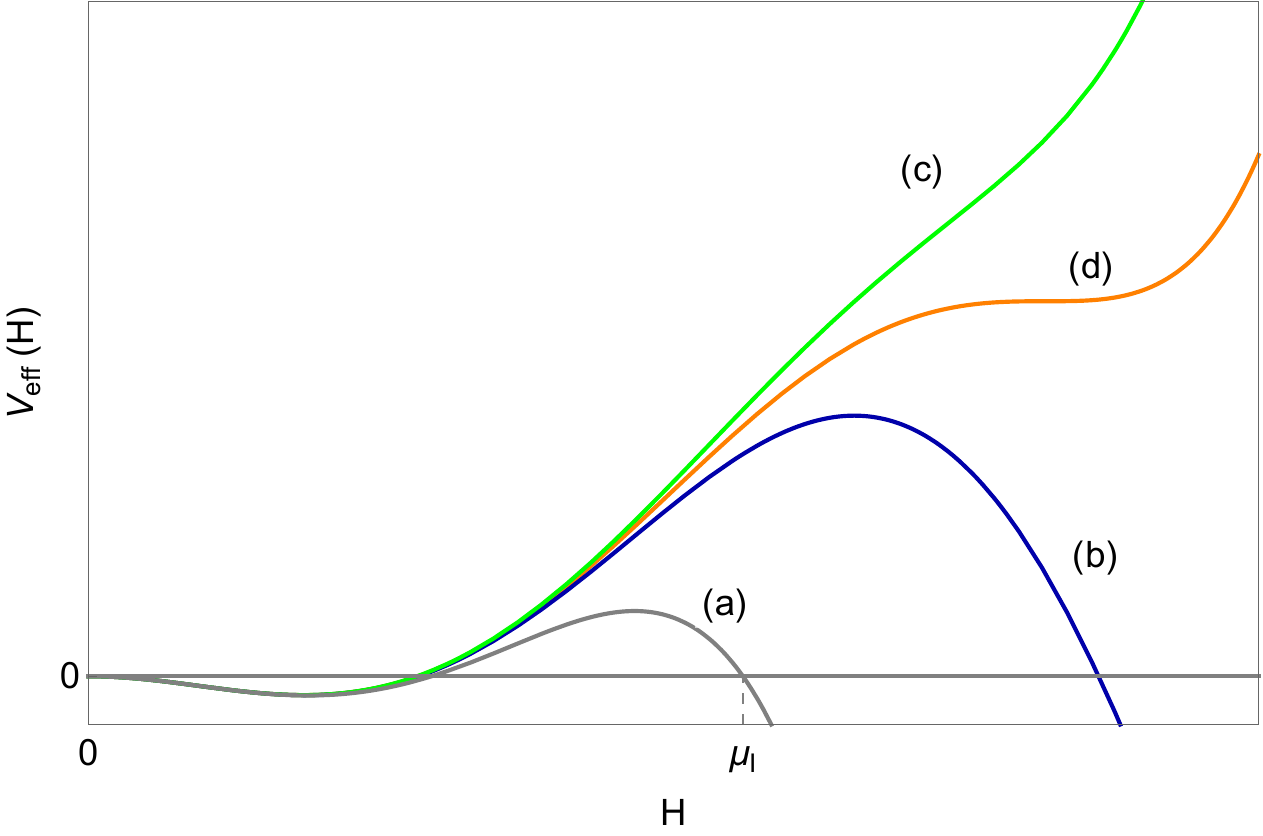}
    \caption{In the absence of a dimension-six term, the effective potential for the Higgs field with a small mass term becomes negative at the instability scale $\mu_I$, at which $\lambda_{\rm eff} = 0$ (curve a, gray). If the effective potential includes a dimension-six term, as in Eq.~(\ref{I/Vexta}), the behavior of the potential depends on the coefficient $C_6/ \Lambda^2$, which may be less than (curve b, blue), greater than (curve c, green), or equal to (curve d, orange) a critical value.
    }
    \label{HierCrit}
\end{figure}

The crucial observation is that, given the observed values of $m_{\rm eff}$ and $\lambda_{\rm eff}$, there is exactly one critical value for the combination $C_6/\Lambda^2$. For large enough values of $C_6 / \Lambda^2$, the dimension-six term becomes dominant at sufficiently low field values that $\lambda_{\rm eff}$ has not yet crossed zero (due to its RG running), thereby yielding a convex potential. For sufficiently small values of $C_6 / \Lambda^2$, on the other hand, the potential still flips over near the instability scale and develops a lower-lying minimum at some even larger energies (if $C_6 >0$) or remains unbounded from below (if $C_6 <0$). These two regimes are separated by the critical value given below in Eq.~\eqref{I/con} and illustrated in Fig.~\ref{HierCrit}.

This behavior is strikingly similar to that of the mass term and quartic coupling, whose critical values correspond to transitions between similar regimes. This motivates the following conjecture: Assume that some mechanism tuning the Higgs field's parameters to values near critical values, marking the transition between qualitatively different classes of potentials, is realized in nature. If this mechanism also influences some yet unknown and suitable UV-extension of the SM, the leading-order correction arising from integrating it out at low energies should lie near the critical value
\begin{equation}
\begin{split}\label{I/con}
    \frac{C_6 (\mu_I)}{\Lambda^2}= r(m^2) \cdot \frac{|\beta_\lambda (\mu_I)|}{\mu_I^2}.
\end{split}
\end{equation}
Here $\mu_I$ denotes the \textit{loop corrected instability scale}, i.e., the scale at which the 1-loop corrected effective quartic coupling vanishes, 
\begin{equation}\label{instscale}
    \lambda_{\rm eff} (\mu_I, H=\mu_I)=0.    
\end{equation}
The term $\beta_\lambda$ governs the running of $\lambda$ under RG flow. The function $r(m^2)$ encodes the coefficient's dependence on the mass term relative to the instability scale. For an approximately vanishing $m^2 / \mu_I^2$, marking the emergence of a non-trivial vacuum near the origin of the potential, it is given by $r(0)= 1 / ( 12 \sqrt{e} )$. Meanwhile, for $m_{\text{max}}^2 = e^{-1}|\beta_\lambda (\mu_I)| \mu_I^2$, the largest value allowing for such a vacuum, we find $r(m_{\text{max}}^2)= \sqrt{e} / 12$.

The remainder of this article is structured as follows. In section~\ref{SMcrit}, we review the near-criticality of the Higgs potential's parameters in greater detail. This includes approaches leading to an alternative critical value for the Higgs mass near $m_{\text{max}}^2$. Sections~\ref{CritPot} and~\ref{NearCritPot} analyze the Higgs potential enhanced by a (near-)critical dimension-six operator, as in Eq.~(\ref{I/Vexta}), and the implications of our results for the applicability of our EFT approach. This includes a discussion of possible observable consequences of such a BSM extension, as well as the relation of our result to the \textit{absolute stability scale}~\cite{Espinosa:2016nld} and the \textit{Multiple Point Principle}~\cite{Froggatt:1995rt,Froggatt:2001pa,Das:2005ku,Froggatt:2005nb,Das:2005eb,Froggatt:2004nn,Froggatt:2006zc,Froggatt:2008hc,Das:2008an,Klinkhamer:2013sos,Froggatt:2014jza,Haba:2014sia,Haba:2014qca,Kawana:2014zxa,Hamada:2015fma,Hamada:2015ria,Haba:2016gqx,Haba:2017quk,McDowall:2018tdg,McDowall:2019knq,Maniatis:2020wfz,Kannike:2020qtw,Kawai:2021lam,Hamada:2014xka,Haba:2014zda,Kawana:2015tka,Hamada:2021jls,Racioppi:2021ynx,Huitu:2022fcw,Hamada:2015dja,Hamada:2017yji,Kawana:2016tkw}. We also consider a recent analysis that predicts the existence of an additional, low-lying de Sitter vacuum below the Planck scale~\cite{Khoury:2022ish}.

In section~\ref{Singlet} we use a simple toy model to illustrate how our results relate to the parameters of concrete BSM models. Lastly, in section~\ref{SOL&MV} we demonstrate how near-critical dimension-six terms naturally relate to a wider class of proposed explanations for the criticality of $m_{\rm eff}$ and $\lambda_{\rm eff}$, in particular self-organized localization~\cite{Giudice:2021viw,Geller:2018xvz} and search optimization on the landscape~\cite{Khoury:2019yoo,Khoury:2019ajl,Kartvelishvili:2020thd,Khoury:2021zao}. 

\section{Near-criticality of the SM Higgs potential}\label{SMcrit}
In the SM, the dynamics of the Higgs field are determined by its potential, which in unitary gauge takes the form
\begin{equation}\label{VSM}
    V_{\rm eff}(H)=- \frac{1}{4} m_{\rm eff}^2 (\mu, H) H^2 + \frac{1}{4} \lambda_{\rm eff}(\mu, H) H^4.
\end{equation}
An astonishing feature of this potential is that both of its parameters appear fine-tuned towards critical values, corresponding to the emergence of new qualitative features.
In the case of the mass term, this behavior leads to the well-known hierarchy problem, whereas the RG trajectory of the quartic coupling is linked to the vacuum's apparent metastability.

Assuming the existence of some more fundamental theory underlying the SM, the potential in Eq.~\eqref{VSM} is to be understood as an effective potential, obtained by integrating out the degrees of freedom heavier than the theory's cutoff $\Lambda$. Besides the introduction of higher-dimensional terms, one would expect that doing so should also induce a \textit{threshold correction} at the matching scale $\mu_{\text{match}}\sim \Lambda$. Assuming the Higgs to be a fundamental scalar, these corrections are generally of the form
\begin{equation}\label{Threshold}
	    \delta m^2 \sim \pm \frac{ \gamma}{(4 \pi)^2} \Lambda^2+ ...,
\end{equation}
with ``+'' for a boson and ``$-$'' for a fermion, and where $\gamma$ represents a suitable combination of couplings.\footnote{Note that in our EFT treatment the corrections induced by the heavy degrees of freedom are absorbed into the mass parameter of the effective theory.} Meanwhile, the absence of new physics in observations near the SM electroweak symmetry-breaking scale $v_{\rm EW}$ suggests either that $\Lambda \gg v_{\rm EW} \sim m$, or that the couplings between the SM and the new physics nearly vanish. Near-vanishing couplings would be consistent with the relation in Eq.~\eqref{Threshold}, but would raise the question of why these couplings are so small in the first place. Meanwhile, for non-vanishing couplings, a small Higgs mass in the IR theory would only be possible if the threshold correction were almost exactly cancelled by the UV mass term, i.e., if the latter were fine-tuned.

A simple way to avoid corrections of the form given in Eq.~\eqref{Threshold} is offered by theories in which the Higgs field is not a fundamental scalar, but rather emerges from some underlying theory, such as in composite Higgs models~\cite{Cacciapaglia:2014uja,Cacciapaglia:2020kgq,BuarqueFranzosi:2018eaj}. However, such theories typically still predict $m^2 \sim (\text{1-loop})\cdot \Lambda^2$. As a consequence, the small observed Higgs mass typically requires some tuning between the parameters of the UV theory, causing an almost-exact cancellation of the coefficient.

Alternatively, the smallness of the mass parameter can be understood as a manifestation of near-criticality, that is, that $m^2$ is close to its critical value $m_{\rm crit}^2=0$. The value $m_{\rm crit}^2$ marks the transition between two phases of the potential: For $m^2 >0$, the potential is convex at the origin, allowing for spontaneous symmetry breaking, whereas for $m^2<0$ the potential is strictly concave and no additional structure is formed. For $m^2=0$, the potential is still strictly convex due to the quartic term, but features a relatively flat region around $H=0$, which may be understood as an example of a \textit{critical feature}. In other words, the hierarchy problem may be rephrased as the question of why the mass parameter lies so close to its critical value. 

The apparent tuning of the quartic coupling~$\lambda$ follows directly from the values of the running couplings near the top-quark mass scale. Its running towards higher energies is determined by its beta function, which, to leading order and neglecting numerical coefficients for now, is of the form
\begin{equation}
    \beta_{\lambda}\propto  \lambda^2 - y_t^4 + \lambda (...)+\text{(gauge terms)}.
\end{equation}
Depending on the relative size of the couplings, this expression gives rise to two distinct phases with a clear transition line. Assuming a small enough Yukawa coupling for the top quark $y_t$, $\lambda$ either grows towards higher energies or declines slowly enough to remain positive until some large energy scale. Since the Yukawa coupling itself runs toward smaller values due to its QCD charge, its negative contribution to $\beta_{\lambda}$ is eventually outmatched by the gauge terms, causing $\lambda$ to run back towards larger values, preventing it from ever becoming negative. This corresponds to a strictly increasing potential, and thus a stable electroweak vacuum.

If, on the other hand, the Yukawa coupling is large enough, its effect can drive $\lambda$ to negative values at a sub-Planckian scale. This behavior makes the Higgs potential ``flip'' and become negative, thus giving rise to regions of field space with energies lower than that of the electroweak vacuum. As a consequence, the field can tunnel into the corresponding part of field space through the nucleation and subsequent expansion of a true vacuum bubble, triggering \textit{electroweak vacuum decay}~\cite{Sher:1988mj,Buttazzo:2013uya,Andreassen:2017rzq}. 

These two classes of potentials may be understood as different phases, determined by the trajectory of $\lambda$ under RG flow, or, equivalently, by the values of the running couplings at some given reference scale $\mu$. Much like the behavior of $V_{\rm eff} (H)$ for values of $m_{\rm eff}$ above or below $m_{\rm crit}$, one of these phases describes a strictly concave potential, while the other contains a turning point and gives rise to a lower-lying, non-trivial vacuum. The critical point can be identified as the value $\lambda_{\rm crit}$ at which the potential develops a saddle point or a second minimum degenerate with the electroweak vacuum. In terms of the RG trajectory of $\lambda$, $\lambda = \lambda_{\rm crit}$ corresponds to $\lambda$ developing a minimum at a sub-Planckian scale, at which it either vanishes or becomes comparable to the logarithmic one-loop terms.\footnote{Due to the high sensitivity of the potential near its critical points, the parameters corresponding to these two scenarios are nearly degenerate.} 

All current observations imply that the relevant couplings lie remarkably close to such a critical point, suggesting that the quartic coupling becomes negative around the instability scale $\mu_I\sim 10^{11}$~GeV, beyond which it remains negative at least until the Planck scale: $-0.01 \leq \lambda (\mu) < 0$ for $\mu_I \leq \mu < M_{\rm pl}$. This implies that the electroweak vacuum can decay, but only at a rate small enough to ensure a lifetime much longer than current age of the universe, $\tau_{\text{SM}}\sim 10^{983}$~years~\cite{Khoury:2021zao,Steingasser:2022yqx}. 

\begin{figure}[t!]
    \centering
    \includegraphics[width=0.45\textwidth]{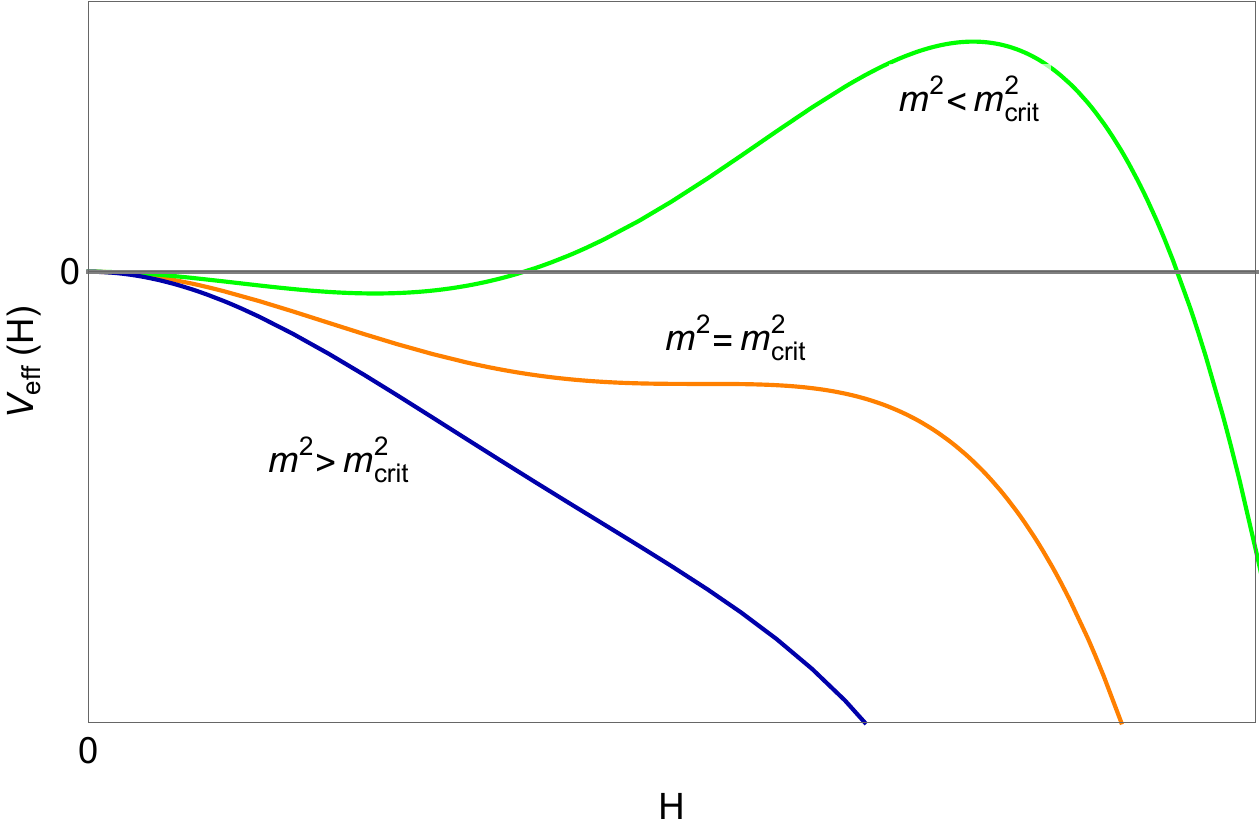}
    \caption{The critical value for the mass term favored by the mechanisms outlined in Refs.~\cite{Giudice:2021viw,Khoury:2019yoo,Khoury:2019ajl,Kartvelishvili:2020thd,Khoury:2021zao}. For $m^2  < m_{\rm crit}^2$, a potential barrier forms (green), whereas $m^2  > m_{\rm crit}^2$ yields a strictly concave potential, with no stable vacua in the IR (blue).}
    \label{MSBound}
\end{figure}

The predicted lifetime of the metastable electroweak vacuum is sensitive to yet-unobserved new physics, since the large value of the instability scale $\mu_I$ allows for many types of BSM fields to affect the running of $\lambda$. For example, if the Higgs field were a composite rather than a fundamental scalar, or if bosonic BSM fields coupled to the Higgs field and thereby modified the running of $\lambda$ to avoid negative values, then $\lambda$ would no longer appear tuned to a near-critical value. On the other hand, fermionic BSM couplings, such as heavy right-handed neutrinos, could drive $\lambda$ to more negative values, lowering the instability scale $\mu_I < 10^{11}$ GeV (perhaps to as low as a few hundred TeV) and yielding a significant destablization of the vacuum~\cite{Chigusa:2018uuj,Khoury:2021zao,Chauhan:2023pur,Giudice:2021viw}. In the latter scenario, the mass term could also be understood as being close to an alternative critical value at
\begin{equation}\label{msbound}
	m^2_{\rm crit}= \frac{|\beta_\lambda|}{e^{\frac{3}{2}}} \mu_I^2 .
\end{equation}
For $m^2 < m_{\rm crit}^2$, the potential would take a form similar to that of the SM, with a non-trivial vacuum at some scale of order $v^2 \sim m^2$ followed by a potential barrier and ultimately a lower-lying vacuum at very high energies. For $m^2  > m_{\rm crit}^2$, the potential would be strictly concave, 
with no minimum below the Planck-scale.\footnote{This insight, together with the observation that such a minimum does indeed exist, implies an upper bound on the Higgs mass for metastable vacua: the metastability bound~\cite{Khoury:2021zao}.} See Fig.~\ref{MSBound}.

\section{Critical extension of the Higgs potential}\label{CritPot}
If both parameters of the SM Higgs potential are indeed tuned toward critical values, it is interesting to consider mechanisms that would yield such near-critical behavior: not only for the parameters within the SM itself (considered as an effective theory), but also within some more fundamental, UV-complete theory.

The most efficient way to investigate this question is to work with an effective theory, in which BSM effects can be described via higher-dimensional operators. In this framework, the Higgs potential takes the form
\begin{align}\label{I/Vext}
 V_{\rm eff} (\mu, H) =& - \frac{1}{4} m_{\rm eff}^2 (\mu, H) H^2 +  \frac{1}{4} \lambda_{\rm eff} (\mu, H) H^4 + \nonumber \\ 
 &+ \frac{ C_6}{\Lambda^2} H^6 +...,
\end{align}
where $\Lambda$ is the scale of new physics and $m_{\rm eff}^2 (\mu, H)$, $\lambda_{\rm eff} (\mu, H)$ are the one-loop-improved mass term and quartic self-coupling respectively. Eq.~\eqref{I/Vext} offers a reliable approximation if the critical feature of the potential occurs at scales $\mu ,H \lesssim \Lambda$. This assumption does not cover every possible scenario, but, as we will see, any BSM model that can be approximated by Eq.~(\ref{I/Vext}) allows for at least one phase transition below the scale $\Lambda$. 

An important subtlety of our argument is the possibility that BSM physics also influences the Higgs field's criticality through corrections to the running of the quartic coupling $\lambda$, for example via momentum-dependent higher-dimensionsl terms in the effective theory. To leading order and in the Warsaw basis~\cite{Grzadkowski:2010es}, this class of operators can be fully characterized through two dimension-six terms,
\begin{equation}
    \mathcal{L}_{D}^{(6)}= \frac{C_{\rm H \Box }}{\Lambda^2} \cdot H^2 \Box (H^2) + \frac{C_{\rm  H D}}{\Lambda^2} \cdot H^2 D_\mu H D^\mu H .
\end{equation}
These terms, as well as all other dimension-six terms including the correction to the potential, affect the running of the quartic coupling through additional terms of the form $\Delta \beta_\lambda \sim C \cdot m^2/\Lambda^2 $.~\cite{Jenkins:2013zja,Jenkins:2013wua,Alonso:2013hga,Alonso:2014zka,Celis:2017hod} Here $m^2$ is the (running) Higgs mass parameter and $C$ represents the Wilson coefficient of the operator of interest. Throughout the remainder of this article, we will find that the scale of new physics lies near or above the instability scale~$\mu_I$, which itself lies multiple orders of magnitude above the Higgs mass. As an immediate consequence, the corrections of the higher-dimensional terms on the running are subdominant in the context of our discussion, ensuring the validity of our analysis at the given level of accuracy. 

\subsection{Critical dimension-six operator}\label{critdimsix}
In order to calculate the critical value $C_6 /\Lambda^2$, we note that the number of distinct phases associated with the potential can, in general, be characterized by the number of local extrema of the potential. The strictly convex phase is characterized by a single minimum (at the origin), whereas in the potentially unstable phase an additional local maximum and minimum emerge. The transition between the two regimes occurs when the local extrema (away from the origin) merge into a single saddle point. Taking into account the minimum induced in a similar way by the mass term, this implies that the critical point is marked by the existence of two local extrema away from the origin. See Fig.~\ref{HierCrit}.

\begin{figure}[t!]
    \centering
    \includegraphics[width=0.45\textwidth]{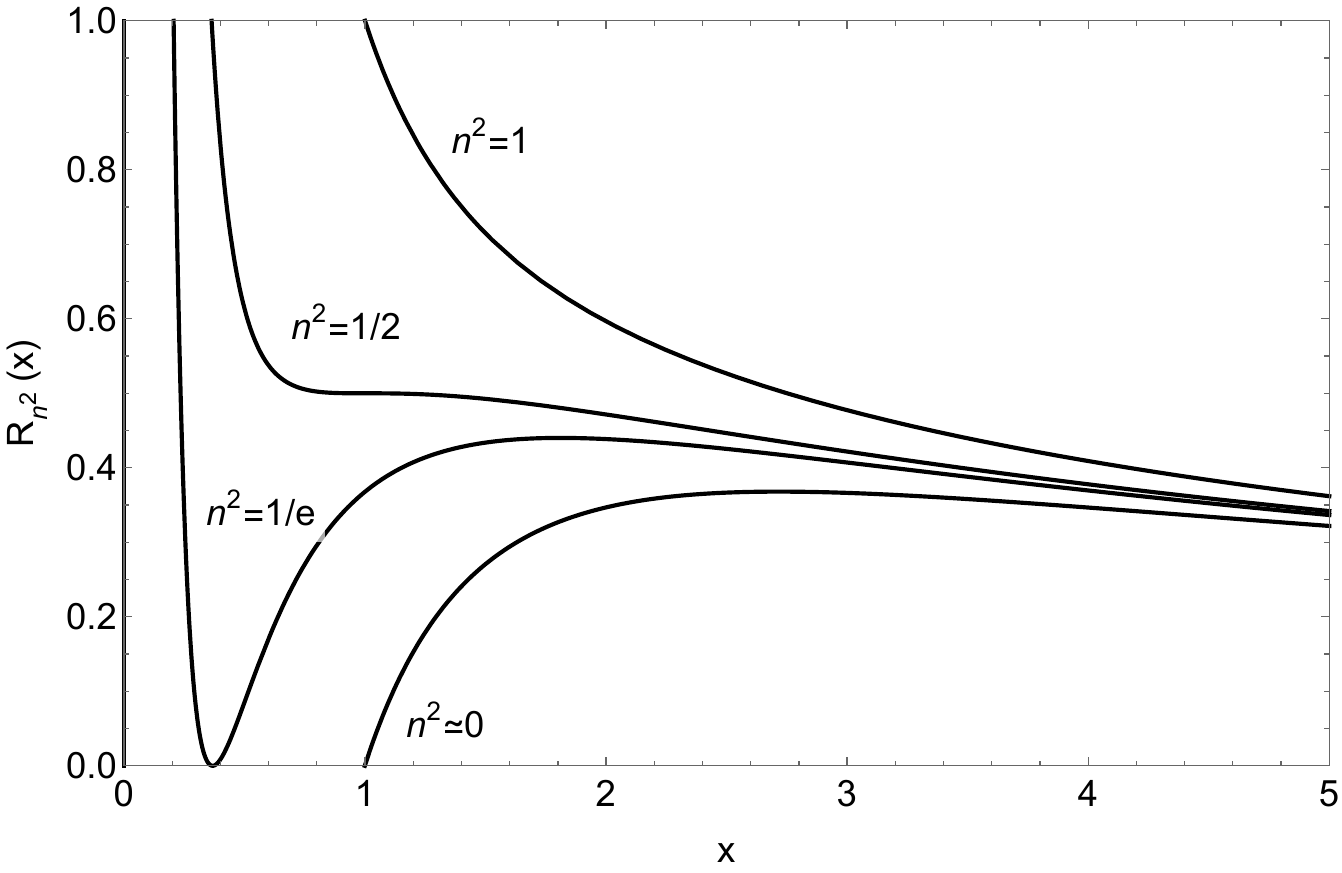}
    \caption{The function~$R_{n^2}(x)$ for different values of $n^2$. The case $n^2=0$ describes a potential with a large hierarchy ($m^2 \ll \mu_I^2 \sim \Lambda^2$), whereas the cases $n^2= 1 / e$ and $n^2 = 1/2$ are related to the metastability bound. The value $n^2=1$ lies beyond the latter, corresponding to a potential with a single minimum near the matching scale.}
    \label{MasterFunction}
\end{figure}

In other words, the potential exhibits critical behavior if there exist precisely two non-zero solutions to the equation
\begin{widetext}
\begin{align}\label{exeq}
  0= \frac{ \partial  V_{\rm eff}(\mu,H) }{\partial H }\bigg\vert_{\mu =H=\bar{H}}=- \frac{m^2_{\rm eff} }{2} \bar{H} -\frac{1}{4} \left( \frac{\partial m^2_{\rm eff} }{\partial H} \right) \bar{H}^2 + \lambda_{\rm eff} \bar{H}^3 + \frac{1}{4} \left( \frac{\partial \lambda_{\rm eff} }{\partial H} \right) \bar{H}^4 + 6 \frac{ C_6 }{\Lambda^2} \bar{H}^5.
\end{align}
\end{widetext}
As the critical point corresponds to a saddle point near the instability scale, we can simplify this equation to leading order by using the following relations:
\begin{gather*}
    m^2_{\rm eff}  \simeq m^2 (\mu_I), \ \ \  \lambda_{\rm eff} \simeq \beta_\lambda (\mu_I) \ln \frac{\bar{H}}{\mu_I},  \ \ \   C_6  \simeq C_6 (\mu_I), \\ 
    \frac{\partial \lambda_{\rm eff}}{\partial H} \simeq \frac{1}{\bar{H}} \beta_\lambda (\mu_I), \ \ \ \frac{\partial m^2_{\rm eff}}{\partial H} = m^2 \cdot \text{(1-loop)}\ll m^2.
\end{gather*}
These allow us to bring Eq.~\eqref{exeq} to the following form:
\begin{equation}\label{eqcompl}
    m^2 = 2 \beta_\lambda \left( \ln \frac{\bar{H}}{\mu_I} + \frac{1}{4} \right) \bar{H}^2 + 12 \frac{C_6}{\Lambda^2} \bar{H}^{4},
\end{equation}
where all parameters are to be understood as evaluated at the instability scale. To make the algebraic structure of this relation more transparent, we may introduce the following set of rescaled parameters:
\begin{gather}
    \mu_t := e^{- \frac{1}{4}} \mu_I  ,  \ \ \ \ 
    n^2 := \frac{m^2}{|\beta_\lambda| \mu_t^2}, \nonumber \\    c:= 12 \frac{C_6}{|\beta_\lambda|} \cdot \frac{\mu_t^2 }{\Lambda^2}, \ \ \ \ x:=\frac{\bar{H}^2}{\mu_t^2}. \label{rescale}
\end{gather}
Thus, we ultimately arrive at the relation
\begin{equation}\label{maseqgen}
    c = \frac{n^2}{x^2}+ \frac{\ln x}{x} =: R_{n^2}(x).
\end{equation}
The function $R_{n^2}(x)$ is shown in Fig.~\ref{MasterFunction} for multiple values of $n^2$, corresponding to different masses. It is easy to see that the solutions of Eq.~\eqref{maseqgen} describe the intersection points of this function with horizontal lines at $y=c$. 

Let us first consider the case $c=0$, which corresponds to $C_6 = 0$ and hence the pure SM case. In this limit, the local extrema can be identified with the intersection points of $R_{n^2}(x)$ with the $x$-axis. For $n^2< 1/e$, i.e., a small Higgs mass relative to the instability scale, we find two such points, corresponding to the electroweak vacuum and the top of the potential barrier. Meanwhile, for $n^2> 1 /e$, there exists no such solution. In terms of the potential, this can be understood as the potential being dominated by the negative mass term for small $H$, while the quartic term becomes dominant only in a regime in which the quartic coupling $\lambda$ has already become negative, such that no minimum exists. The transition between these two phases is given by the critical point $n^2 = 1/e$, for which the electroweak vacuum and the potential barrier merge into a saddle point. Thus, demanding the existence of the electroweak vacuum implies an upper bound on the mass parameter, known as the metastability bound~\cite{Khoury:2021zao,Steingasser:2022yqx}. 
Based on the simple picture given in Fig.~\ref{MasterFunction}, it is easy to see that a similar bound can be recovered for a non-vanishing dimension-six term. Due to its sign, the latter contributes to the formation of the electroweak vacuum, thus allowing for a larger mass term and thereby weakening the bound from $n^2 < 1/e$ up to $n^2 < 1/2$. For larger values, there exists only one minimum at which the negative quadratic and quartic terms become comparable to the dimension-six term. For smaller values, one recovers the two extrema of the SM-like potential and an additional (typically lower-lying) minimum at some larger value of the Higgs field at which the dimension-six term becomes dominant.

\begin{figure}[t!]
    \centering
    \includegraphics[width=0.45\textwidth]{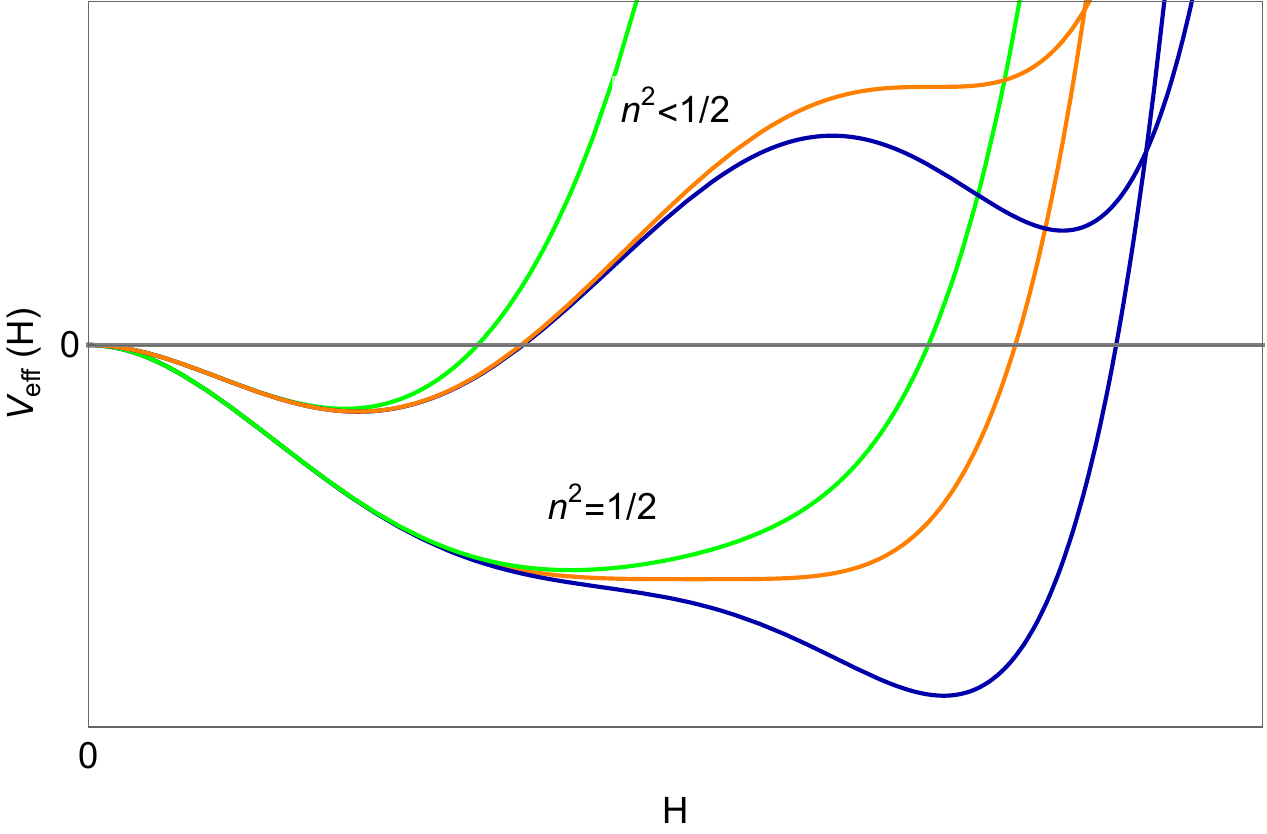}
    \caption{The effective potential for various near-critical values of the dimension-six operator.\newline
    $n^2= 1 / 2$: Saturation of the generalized metastability bound implies a unique minimum for all non-vanishing dimension-six terms. For the critical value this minimum is flat $V''=0$ (orange), while smaller (blue) as well as larger values (green) differ from one another by the existence of a turning point. \newline
    $n^2< 1 / 2$: Below the metastability bound the potential has the same qualitative features as for a large hierarchy (that is, for $m^2 \ll \mu_I^2\sim \Lambda^2$). The critical dimension-six term thus corresponds to a saddle-point (orange), separating the phases generated by smaller (blue) and larger (green) values.
    }
    \label{BoundCrit}
\end{figure}

On the other hand, for any given value of $n^2$, the number of local extrema depends on the value of $c$. Assuming $n^2 < 1/2$, let $c_{n, \rm crit}$ denote the local maximum of $R_{n^2}$. Then, for $c>c_{n, \rm crit}$, there exists only one minimum, as the sign flip of the quartic term occurs in a regime in which the dimension-six term is already dominant. For smaller values of $c$, we again find three extrema, corresponding to the electroweak vacuum, the top of the potential barrier, and a second minimum at the value of $H$ where the dimension-six term becomes dominant. Thus, in terms of our dimensionless quantities, the critical point is given by $c_{n, \rm crit}$, which represents two extrema: the electroweak vacuum as well as a saddle point near the instability scale. See Fig.~\ref{BoundCrit}.

Returning to conventional units, the critical value of the coefficient $C_6/\Lambda^2$ is given by
\begin{align}\label{C6crit}
    \left( \frac{C_6}{\Lambda^2} \right)_{\rm crit} =  \frac{\sqrt{e}}{12} \cdot c_{n^2,\rm crit} \cdot \frac{|\beta_\lambda|}{\mu_I^2}.
\end{align}
For the two most interesting values of $n^2$, corresponding to a large hierarchy ($n^2 = 0$) and to a value almost saturating the metastability bound ($n^2 \lesssim 1/2$), the coefficient $c_{n^2, \rm crit}$ is easily determined analytically:
\begin{align}
     c_{0, \rm crit}=&\frac{1}{e} \ \ \text{for} \ n^2=0 \ \ (m^2 \ll \mu_I^2 \sim \Lambda^2 ) , \label{ccrit0}\\ 
    c_{1/2, \rm crit}=&\frac{1}{2} \ \ \text{for} \ n^2\lesssim \frac{1}{2} \ \ (m^2\lesssim e^{- \frac{3}{2}} |\beta_\lambda| \mu_I^2) \label{ccrithalf} .
\end{align}
By construction, the critical value in Eq.~\eqref{C6crit} gives rise to a saddle point. Using Eq.~\eqref{C6crit}, it is straightforward to find that the potential at its saddle point $\bar{H}_{n^2}^2$ satisfies
\begin{gather}
    V_{0}(\bar{H}_0)= \frac{e}{48}  |\beta_{\lambda}| \mu_I^4  \ \ \text{at} \ \ \bar{H}_0^2= \sqrt{e} \mu_I^2 = \frac{1}{12} \frac{|\beta_\lambda|}{C_6} \Lambda^2  \label{H0n0} \\ 
    V_{\frac{1}{2}}(\bar{H}_{\frac{1}{2}})= -\frac{|\beta_{\lambda}| }{48 e} \mu_I^4  \ \ \text{at} \ \ \bar{H}_{\frac{1}{2}}^2= \frac{\mu_I^2}{\sqrt{e}} = \frac{1}{24} \frac{|\beta_\lambda|}{C_6} \Lambda^2 . \label{H0nhalf}
\end{gather}
Note from Eqs.~(\ref{C6crit})--(\ref{ccrithalf}) that the critical value of $C_6 / \Lambda^2$ itself depends on $n^2$, which is incorporated into our expressions for $\bar{H}^2_{n^2}$ in Eqs.~(\ref{H0n0})--(\ref{H0nhalf}). 

The defining feature of the critical dimension-six term, the formation of a saddle point, also immediately implies the gauge invariance of our result through the Nielson identity~\cite{Nielsen:1975fs,Fukuda:1975di}.\footnote{This feature can also be understood directly from Eq.~\eqref{C6crit}: The running of the $\bar{\text{ms}}$-coupling $\lambda$ is gauge-independent, and so is $\beta_{\lambda}$. It was furthermore shown that the gauge-dependence of its loop-corrections vanish at the one-loop level near the instability scale~\cite{Andreassen:2014gha}. Together, these two observations imply that $\mu_I$, as defined in Eq.~\eqref{instscale}, is indeed gauge-invariant at the order of interest.}

\subsection{Implications of criticality}\label{CritImp}
Eq.~\eqref{C6crit} relates the scale of new (bosonic) physics, represented through $\Lambda$ and $C_6$, to the instability scale $\mu_I$ as well as the running of the quartic coupling in its vicinity, described by $\beta_\lambda$. Although the exact interpretation of this relation depends on the details of the mechanism ultimately responsible for the near-criticality and the choice of free parameters, it is nevertheless possible to anticipate some of its aspects.

We may first note that Eq.~\eqref{C6crit} is independent of which parameters are assumed to be subject to an underlying dynamical mechanism. As the presence of \textit{some} dimension-six term is arguably less surprising than the vanishing of the quartic coupling, one possible interpretation of our result is that the parameter sensitive to dynamical selection is the quartic coupling at the matching scale, $\lambda_{\rm eff} (\mu_I , \mu_I)$, while the dimension-six term is either a relic of the physics underlying the selection mechanism or some unrelated physics, such as a spectator field or even quantum gravity.\footnote{We will explore the range of relevant scales for $\Lambda$ in section~\ref{C6crit}, where we find that the Planck scale lies indeed within the 95\%-confidence interval.} This perspective would suggest that the critical behavior described by Eq.~(\ref{C6crit}) is not an intrinsic property of the dimension-six term, but rather determines the scale at which the critical feature of the quartic coupling manifests.

An alternative approach would be to understand Eq.~\eqref{C6crit} as a prediction for the dimension-six term and treat the RG trajectory of $\lambda$ (governed by $\beta_\lambda$) as given. This is not necessarily favored by any theoretical arguments, but is nevertheless the more convenient choice for the purpose of making concrete predictions: the quartic coupling has been inferred with great accuracy in the vicinity of the electroweak symmetry-breaking scale, and its running under RG flow is well-understood, even taking into account the effects of yet unknown BSM physics at scales below $\Lambda$. We will therefore adopt this perspective for the remainder of this article.

Next we note that Eqs.~\eqref{H0n0}-\eqref{H0nhalf} imply that the critical feature in the effective potential, separating distinct phases, emerges at field values at or slightly above the instability scale $\mu_I$, independent of details of the UV theory. Considering $\beta_\lambda$ as given, these equations also imply a simple relation between the Wilson coefficient $C_6$ and the scale of new physics $\Lambda$. First, large $C_6$ imply a relatively large $\Lambda$, typically ensuring the validity of our EFT approach. Conversely, large values of $\Lambda$ relative to the instability scale are only possible for relatively large values of $C_6$. If the coefficient $C_6$ is entirely loop-induced, such that $C_6$ is a function of other couplings, then large values of $C_6$ would violate perturbativity in the UV theory. As a consequence, criticality of the dimension-six term implies an upper bound on the scale $\Lambda$ of new physics for such theories.

Meanwhile, small values of $C_6$ imply the emergence of the critical feature beyond the self-consistent regime of the effective theory, where the logarithmic corrections manifesting in the dimension-six term would typically be absorbed into the RG-running. We discuss an explicit example for such a case in section~\ref{Singlet}, where we illustrate that this indeed prevents the formation of a critical feature in the sense we discussed in this article, rather suggesting the notion of criticality proposed in Refs.~\cite{Bezrukov:2014bra,Hamada:2014wna,Iacobellis:2016eof,Enckell:2016xse,Rasanen:2017ivk,Enckell:2020lvn,Salvio:2018rv,Lee:2023wdm,Kawai:2023vjo}.

\subsection{Range of scales}\label{rangeofscales}

Eq.~(\ref{C6crit}) implies that the critical value of $C_6 /\Lambda^2$ is given, up to numerical coefficients, by the one-loop instability scale $\mu_I$. Assuming $C_6 \sim {\cal O} (1)$, our conjecture then suggests that the scale $\Lambda$ of new (bosonic) physics should roughly coincide with the instability scale. The value of the latter is therefore not only crucial for the search for concrete realizations of our conjecture, but also for the possibility of experimental tests of our conjecture, and by extension the more fundamental idea of some BSM physics favoring near-critical behavior.

\begin{figure}[t!]
    \centering
    \includegraphics[width=0.45\textwidth]{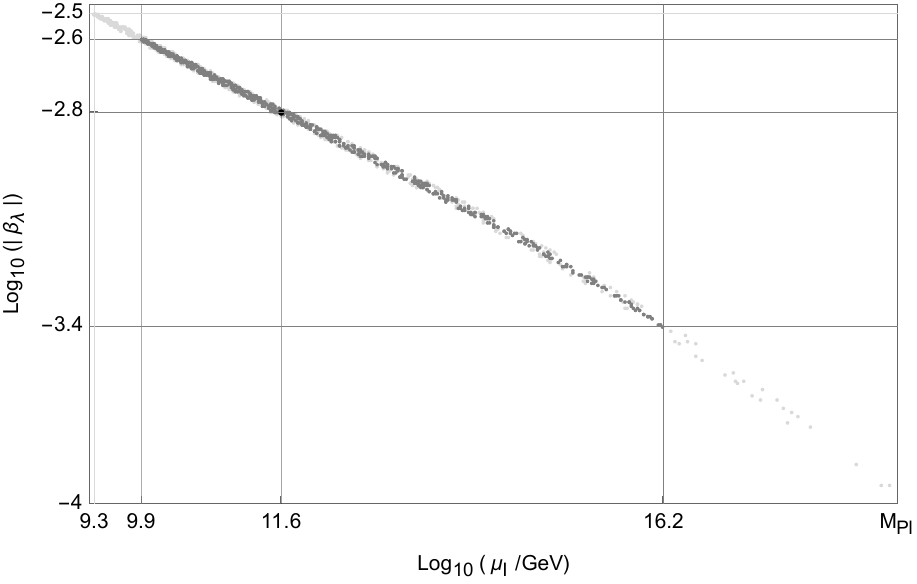}
    \caption{The phenomenologically favored range of the two IR parameters determining the properties of the critical point, $\beta_{\lambda}(\mu_I)$ and $\mu_I$, obtained by integrating the SM $\beta$ functions at three-loops, starting from the IR matching values in Eq.~(\ref{gmatching}). The central value (black) is $\mu_I \sim 4 \cdot 10^{11}$~GeV and $\beta_{\lambda}(\mu_I)=1.65\cdot 10^{-3}$. The dark and light gray regions represent the 68\% and 95\% confidence ellipses, which we obtained by scanning over the corresponding ellispoids in $\{\lambda (M_t),y(M_t),g_s(M_t)\}$-space.}
    \label{MuIBeta}
\end{figure}

The value of the instability scale $\mu_I$ is primarily determined by the RG trajectory of $\lambda$, with some corrections coming from its loop-corrections. It can therefore be easily determined by integrating the SM $\beta$ functions, which can be found, e.g. in the appendices of Refs.~\cite{Buttazzo:2013uya,Khoury:2021zao,Chauhan:2023pur}, starting from the measured values of the relevant couplings at the top-quark mass scale $M_t$~\cite{Huang:2020hdv}, 
\begin{gather}
    \lambda (M_t)=0.12607 \pm 0.00030 , \  y_t (M_t)=0.9312 \pm 0.0022,  \nonumber \\ 
     g(M_t)=0.64765 \pm  0.00028 , \ g_s (M_t)=1.1618 \pm 0.0045,  \nonumber \\  g^\prime (M_t)=0.358545 \pm 0.000070.   \label{gmatching}
\end{gather}
The couplings' error bars are a direct result of observational uncertainties of the masses of the top quark and the Higgs boson as well as the QCD coupling, from which they were obtained by a combination of linear error propagation and Monte Carlo simulations by the authors of Ref.~\cite{Huang:2020hdv}. This allows for a straightforward construction of error ellipsoids in the space of the couplings at the IR matching scale $M_t$. By scanning over a lattice of 718 points on their surfaces each for $68\%$ and $95\%$ confidence, we ultimately arrive at the following ranges for the one-loop improved instability scale (with three-loop accuracy in the running) as well as the scale $\mu_I$:
\begin{gather}
    8 \cdot 10^9~\text{GeV} < \mu_{I,68\%} < 10^{16}~\text{GeV},  \\  3 \cdot 10^9~\text{GeV} < \mu_{I,95\%}\label{murange}
\end{gather}
These values are illustrated in Fig.~\ref{MuIBeta}. Note that the near-criticality of the RG trajectory of $\lambda$ also manifests in the error bar. The critical behavior corresponds to a positive quartic coupling for all scales, such that approaching it leads to a quickly growing $\mu_I$. We find in particular that the error ellipsoid corresponding to a 95\% confidence level also contains a subset of points corresponding to absolutely stable vacua. See Fig.~\ref{MuIBeta}.

Assuming that all relevant BSM effects on the Higgs effective potential (up to scale $\Lambda$) are represented by the $C_6$ term in Eq.~(\ref{I/Vext}), and given the relation $\mu_I \sim \Lambda$, the expected scale for new physics lies many orders of magnitude beyond existing and anticipated experiments. Hence our conjecture of near-criticality for the Higgs sector could easily be contradicted if new bosonic degrees of freedom are detected with masses below $\mu_I$ and with couplings to the Higgs field of nontrivial strength. On the other hand, if such degrees of freedom remain absent even in next-generation experiments, the hypothesis of near-criticality for $V_{\rm eff} (\mu, H)$ may provide a useful basis for BSM model-building. Hence the lower end of the interval in Eq.~(\ref{murange}), $\mu_I \gtrsim {\cal O} (10^9)$ GeV, is of particular interest in the context of direct-detection efforts. Following our previous discussion, this provides a lower bound on the scale of new physics in general agreement with our conjecture.

We note further that the instability scale $\mu_I$ is sensitive to new physics. In particular, BSM fermions, such as right-handed neutrinos, could significantly lower $\mu_I$, since they typically introduce a negative contribution to $\beta_\lambda$. In the absence of a $C_6 H^6 / \Lambda^2$ term in $V_{\rm eff}$, such BSM fermions would significantly destablize the electroweak vacuum. In the presence of a $C_6 H^6 / \Lambda^2$ term, however, such BSM fermions could shift $\mu_I$ as low as the mass of the lightest additional fermion {\it without} triggering vacuum instability (given the stabilizing effect of the additional $H^6$ term on $V_{\rm eff}$). As concrete and well-motivated examples, right-handed neutrinos involved in a low-scale seesaw mechanism and vector-like quarks could yield $\mu_I \sim$ TeV~\cite{Khoury:2021zao,Chauhan:2023pur,Belfatto:2023tbv}. If evidence for new fermions emerges, then in the context of our near-criticality proposal, that would suggest that new bosonic physics should {\it also} emerge at energy scales about one order of magnitude greater.\footnote{It is important to note that this second conclusion would be true independent of our conjecture, as such large couplings would require some sort of stabilization. The special role our conjecture plays in this context is discussed further in section~\ref{NearCritEx}.}

On the other end of the range for $\mu_I$ in Eq.~(\ref{murange}), we observe that the 95\% confidence interval contains the Planck scale. At this end of the range for $\mu_I \sim \Lambda$, we would not expect to find evidence of new physics via direct-detection experiments, though the critical features of $V_{\rm eff} (\mu, H)$ could have important implications for early-universe models such as Higgs inflation \cite{Bezrukov:2007ep,Bezrukov:2010jz,Bezrukov:2012sa,Bezrukov:2014bra,Bezrukov:2014ipa,Kaiser:1994vs,Greenwood:2012aj,Kaiser:2013sna}. For the central values, the feature emerges at energy scales typically reached during reheating~\cite{Amin:2014eta,Allahverdi:2020bys}, where they could trigger bubble nucleation events. We explore such cosmological scenarios in future work.

\section{Near-critical extension of the Higgs potential}\label{NearCritPot}

In section~\ref{CritPot}, we have found that the critical value of the dimension-six term corresponds to a potential with a saddle point at $\bar{H}_{n^2}$, as given in Eqs.~\eqref{H0n0}--\eqref{H0nhalf} for $n^2= \frac{1}{2}$ and $n^2= 0$. As, however, both the Higgs field's mass term and quartic coupling are only near-critical, as is to be expected if they were the result of some dynamical selection mechanism, our analogy-based reasoning would suggest that the same can be expected for the dimension-six term. In other words, we would expect that the parameter $C_6 / \Lambda^2$ takes values close to its critical value, within one of the two phases depicted in Fig.~\ref{BoundCrit}.

Let us first consider the case $C_6 / \Lambda^2 \gtrsim \left( C_6 / \Lambda^2 \right)_{\rm crit}$ as in the green curves in Fig.~\ref{BoundCrit}, which correspond to a strictly convex potential (besides the behavior near the electroweak scale). This would render the electroweak vacuum absolutely stable, and more generally erase the qualitative features in the potential linked to the critical point. This scenario would still allow an interpretation of the quartic coupling as near-critical in the sense that its value at the Planck scale lies close to the conformal value $\lambda=0$, which is, however, not obviously correlated to the notion used in our initial arguments: Both the mass term and the quartic coupling are tuned precisely such that their corresponding feature manifests, but just barely so. We therefore focus instead on the scenario $C_6 / \Lambda^2 \lesssim \left( C_6 / \Lambda^2 \right)_{\rm crit}$, as in the blue curves in Fig.~\ref{BoundCrit}. 

While subdominant in the context of particle physics, the existence of a second minimum at higher energies can be crucial for the possibility of observational consequences on cosmological scales. First, as pointed out in the previous subsection, it remains possible that the near-critical features appear at scales just slightly below the Planck scale, where they could have a significant impact on the possibility of Higgs inflation~\cite{Kaiser:1994vs,Kaiser:2013sna,Greenwood:2012aj,Bezrukov:2007ep,Bezrukov:2010jz,Bezrukov:2012sa,Bezrukov:2014ipa}. Meanwhile, the central values of the couplings suggest that the second minimum lies within the range of reasonable reheating temperatures~\cite{Amin:2014eta,Allahverdi:2020bys}. This allows for its occupation through thermal fluctuations in the early universe, and, more importantly, its subsequent decay through a first-order phase transition, causing the emission of gravitational waves.\footnote{We will explore both of these possibilities in future works.}

\subsection{The near-critical potential}

The case $C_6 / \Lambda^2 \lesssim \left( C_6 / \Lambda^2 \right)_{\rm crit}$ corresponds to the emergence of a local minimum. The deviation from the critical point can be quantified through the parameter 
\begin{equation}\label{dcdef}
    \epsilon^2=c_{n^2, \rm crit} - c_{n^2},
\end{equation}
which for near-critical potentials can be used as a perturbative expansion parameter. Recall from Eq.~(\ref{rescale}) that $c =  12 e^{-1/2} \vert \beta_\lambda \vert^{-1} (C_6 / \Lambda^2)$.

\begin{figure*}[t!]
    \centering
    \includegraphics[width=0.4\textwidth]{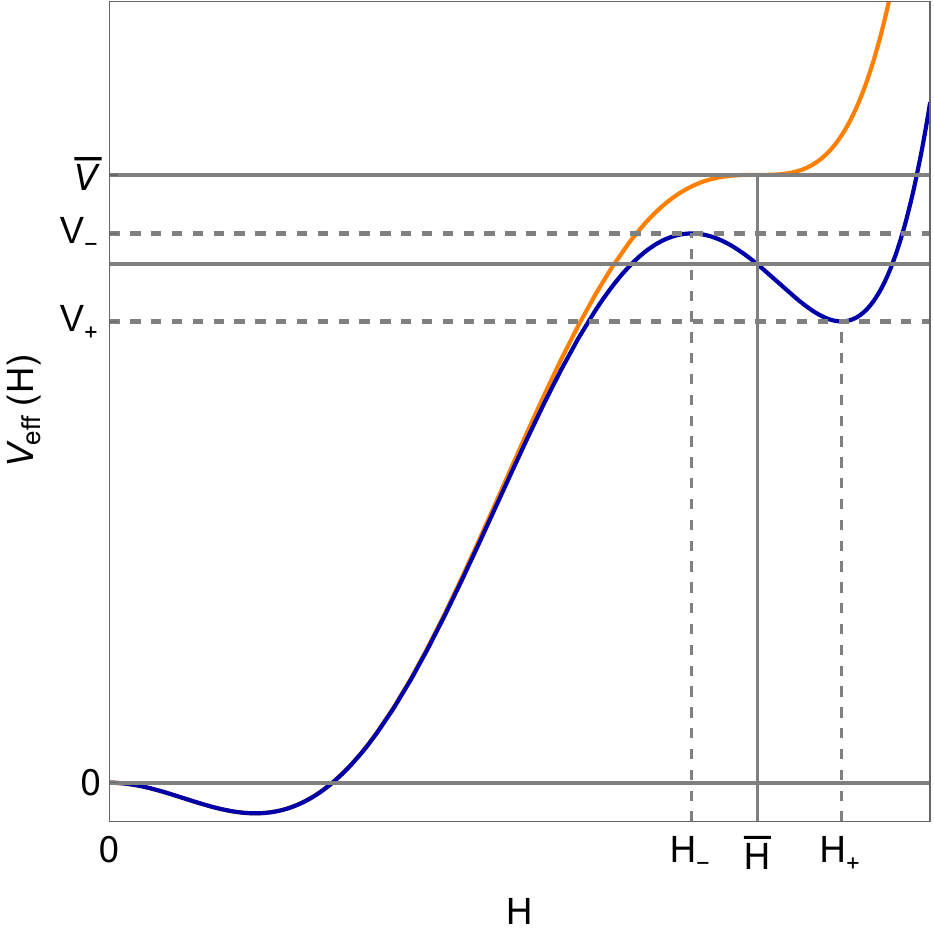}
    \includegraphics[width=0.4\textwidth]{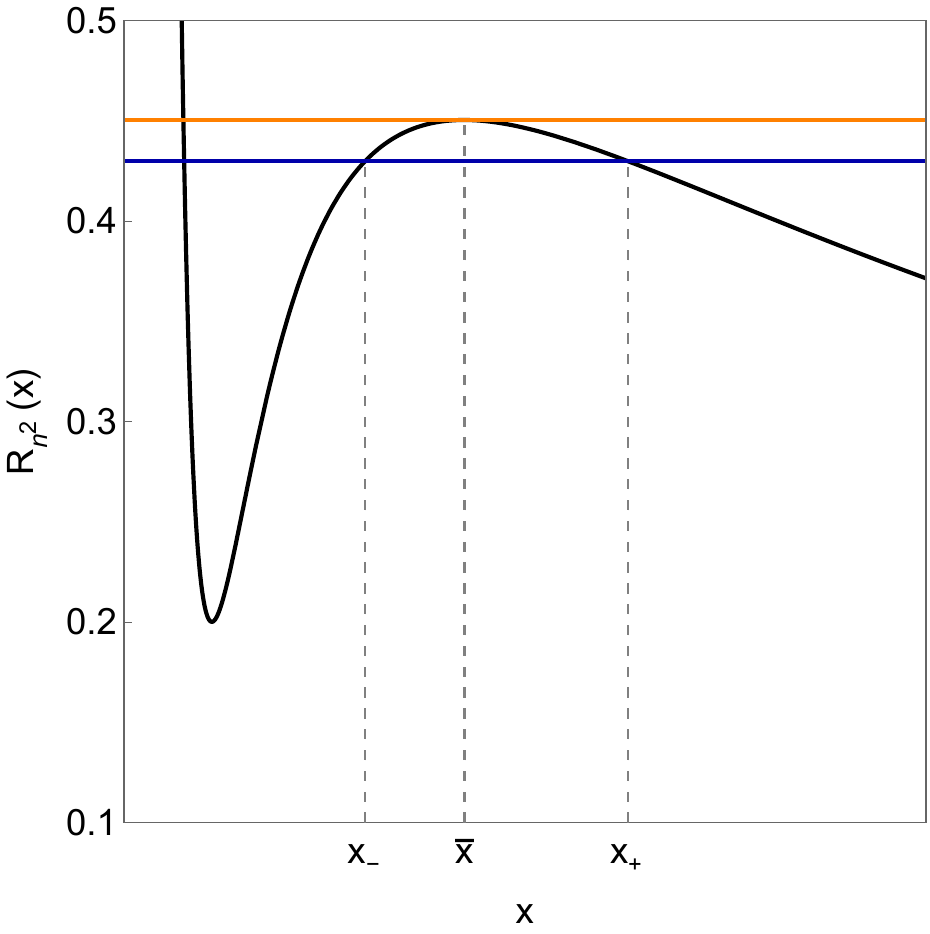}
    \caption{Left: The effective potential with a critical (orange) and near-critical (blue) dimension-six term. $\bar{H}$ and $\bar{V}$ represent the saddle point of the critical potential, as given in Eq.~\eqref{H0n0}--\eqref{H0nhalf}. Considering instead a near-critical dimension-six term, this feature is replaced by a local minimum and maximum at $H_-$, $H_+$, with potential energy $V_-$, $V_+$. See Eq.~\eqref{dHnearcrit}--\eqref{dVnearcrit}. This splitting arises on top of an overall downward shift of the potential, whose leading-order effect is represented by the lower horizontal line. Right: The corresponding equation visualized using the function $R_{n^2}(x)$. $x_{\pm}$ and $\bar{x}$ correspond to $H_{\pm}^2$ and $\bar{H}^2$, respectively.}
    \label{fig:Vcritical}
\end{figure*}

We will first consider the case $n^2 \simeq 0$, corresponding to $m^2 \ll \mu_I^2\sim \Lambda^2$. Assuming $\delta \epsilon >0$ in the sign convention used in Eq.~\eqref{dcdef}, this leads to a potential with two solutions to Eq.~(\ref{eqcompl}), corresponding to a local maximum and minimum. As they replace a saddle point in the critical potential, the splitting in their energies emerges only at next-to-leading order in the perturbative expansion. Meanwhile, the leading order correction manifests through the change of the dimension-six term, $\delta V \sim - \epsilon^2 \cdot \bar{H}_0^2 /\Lambda^2$, which amounts to a lowering of the region around the saddle-point as a whole. Taking into account next-to-leading order corrections in order to resolve the splitting of the extrema, their positions and corresponding energies as depicted in Fig.~\ref{fig:Vcritical} are given by
\begin{gather}
    H_{\pm}^2= \bar{H}^2( 1 \pm  \sqrt{2 e} \cdot \epsilon  + \frac{5}{3} e \cdot  \epsilon^2 + ... ) ,\label{dHnearcrit}
    \\
    V_{\pm}= \bar{V} (1 - 4 e \cdot \epsilon^2 \mp 8 \sqrt{2} e^{\frac{3}{2}} \cdot \epsilon^3+ ... ) .\label{dVnearcrit}
\end{gather}
An important subtlety of this perturbative approach is its limited range of applicability due to the high sensitivity of the potential near its local extrema, arising from its near-criticality. These effects are distinct already for a relatively minor deviation from the critical value in Eq.~\eqref{C6crit}, such that a reliable perturbative treatment requires to take into account relatively high orders in $\epsilon$. On the other hand, this also implies that Eq.~\eqref{C6crit} offers a reliable order-of-magnitude estimate for the entire range of near-critical values. 

The case $n^2\simeq \frac{1}{2}$, corresponding to a near-saturation of the metastability bound~\eqref{msbound}, shows some qualitative differences compared to the one of a large hierarchy $n^2 \simeq 0$. Due to the properties of this critical point, simply perturbing the dimensionless parameter $c$ when $n^2=\frac{1}{2}$ holds exactly amounts to a shift of the still unique minimum of the potential, see Fig.~\ref{BoundCrit}. We can thus focus on the leading order in $\epsilon$, where we find
\begin{align}
    \tilde{H}^2=\bar{H}^2 \left( 1+ \left(6 \epsilon^2 \right)^{\frac{1}{3}} + ... \right), \ \ \ \tilde{V}=\bar{V}\left( 1 + 4 \epsilon^2 + ... \right).
\end{align}
That no second vacuum forms is, however, unique to the critical value $n^2=\frac{1}{2}$. It is easy to see that for slightly smaller masses, $n^2= \frac{1}{2}-\delta n^2$, the familiar structure emerges as long as $\epsilon \gtrsim \delta n$, in principle allowing for an extended perturbative treatment.

\subsection{Examples for near-criticality}\label{NearCritEx}
In the absence of new physics such as the dimension-six term, the electroweak vacuum is most likely metastable, i.e., it can decay by tunneling through the potential barrier. The walls of the resulting bubble carry a positive gradient energy density, which needs to be compensated for by the negative potential energy of the field in its center. As an immediate consequence, this channel of vacuum decay becomes impossible if the potential's two minima are precisely degenerate. 

For a potential of the form in Eq.~\eqref{I/Vext} with a small mass term $m^2 \ll \mu_I^2$, whether or not this is the case can be understood as a condition on the dimension-six term. This motivated the introduction of the \textit{absolute stability scale} $\Lambda_{\text{abs}}$ in Ref.~\cite{Espinosa:2016nld}, which is defined as the value of $\Lambda$ for which the potential's minima are degenerate, assuming the remaining parameters to be given and setting $C_6=1$. Within our framework, the latter can be understood as a special case of a near-critical dimension-six term. The perturbative result in Eq.~\eqref{dVnearcrit} implies that the latter should roughly correspond to $\epsilon^2 \sim 0.3$, clearly outside of the range of its applicability. It is, however, trivial to obtain a full numerical value 
\begin{equation}
    \epsilon^2_{\text{abs}}\simeq 0.033,
\end{equation}
demonstrating the typical sensitivity near a critical point.

A special case of such behavior is the \textit{Multiple Point Principle}~\cite{Froggatt:1995rt,Froggatt:2001pa,Das:2005ku,Froggatt:2005nb,Das:2005eb,Froggatt:2004nn,Froggatt:2006zc,Froggatt:2008hc,Das:2008an,Klinkhamer:2013sos,Froggatt:2014jza,Haba:2014sia,Haba:2014qca,Kawana:2014zxa,Hamada:2015fma,Hamada:2015ria,Haba:2016gqx,Haba:2017quk,McDowall:2018tdg,McDowall:2019knq,Maniatis:2020wfz,Kannike:2020qtw,Kawai:2021lam,Hamada:2014xka,Haba:2014zda,Kawana:2015tka,Hamada:2021jls,Racioppi:2021ynx,Huitu:2022fcw,Hamada:2015dja,Hamada:2017yji,Kawana:2016tkw}, which describes the conjecture that there should exist a second minimum of the Higgs potential degenerate with the electroweak vacuum. Assuming only the SM particle content, this is naturally achieved in the conformal scenario in which $\lambda_{\rm eff} (M_{\text{pl}}) = \beta_\lambda (M_{\text{pl}})=0$, which lies at the boundary of the $95\%$-confidence ellipse. (If Planck-suppressed quantum-gravitational corrections are incorporated, the degeneracy of the vacua would require small negative values of $\lambda (M_{\text{pl}})$ and $\beta_\lambda (M_{\text{pl}})$ near the critical feature.)

Nearly-degenerate vacua are of particular interest in the context of recent results concerning cosmological vacuum selection. It was argued in Ref.~\cite{Khoury:2022ish} that Bayesian arguments favor potentials with a tower of near-degenerate vacua converging towards the critical point marking the transition from de Sitter (dS) to anti-de Sitter (AdS) spacetimes. Setting out from the observed values of the electroweak scale and the cosmological constant, this corresponds to the existence to a series of additional, near-degenerate vacua with vacuum energies $0<V_0<v_{\rm EW}$ below the Planck scale.  Assuming the lowest-lying of these minima to be the result of the dimension-six term, the corresponding value of $\Lambda$ is essentially identical with the absolute stability scale, due to the high sensitivity near critical values.

\section{Explicit realization in a singlet extension}\label{Singlet}

To illustrate the main points of our work, we apply it to the simplest suitable SM extension, that is, the addition of a scalar singlet $S$ with mass $M \sim \Lambda \gg m$. We parametrize the potential of this theory as
\begin{align}\label{VUV}
    V_{\rm UV}(H,S)=&\frac{1}{2}M^2 S^2 + \frac{1}{2}m_H^2 H^2 + \frac{1}{2} \kappa H^2 S^2 + \nonumber \\ 
    +&\frac{1}{4} \lambda_H H^4 + \frac{1}{4} \lambda_S S^4 + \nu H^2 S,
\end{align}
where we adopt the normalization conventions of Ref.~\cite{Jiang:2018pbd} to allow easy comparison with prior literature.

For simplicity, we assume all parameters in this potential to be positive, which implies in particular that $\langle S \rangle =0$ in the absence of a background Higgs field. To further simplify our discussion, we will focus exclusively on the critical point corresponding to $m_{\rm IR}^2=0$ for the remainder of this section and drop the index $n^2$. Upon integrating out the heavy field $S$ at scales smaller than $M$, this implies that the mass parameter for the Higgs field as well as the coefficient of the induced dimension-six term are given at the matching scale by
\begin{gather}
    m_{\rm IR,eff}^2=0,\ \ \
    C_6 = \frac{M^2}{\mu_I^2} \frac{|\beta_\lambda |}{12\sqrt{e}}=\frac{M^2}{\bar{H}^2} \frac{|\beta_\lambda |}{12}.
\end{gather}
Consistent with our previous assumptions, we will furthermore assume a small enough separation between the instability scale $\mu_I$ and the matching scale $M$ to approximate the quartic coupling as
\begin{equation}
    \lambda_{\rm IR,eff}^{\text{(1-loop)}} (M) \simeq - | \beta_\lambda | \frac{1}{2} \ln \frac{M^2}{\mu_I^2} = - | \beta_\lambda | \left( \frac{1}{2} \ln \frac{M^2}{\bar{H}^2} + \frac{1}{4}\right).
\end{equation}
In section~\ref{ExplicitTree}, we first consider the most general case $\nu \neq 0$. This amounts to the generation of a dimension-six term at tree-level, which, following our arguments in section~\ref{CritImp}, is the least restrictive in terms of suitable scales of new physics. In section~\ref{ExplicitLoop}, we then consider the more restricted and hence predictive scenario in which the dimension-six term is entirely induced by loop-corrections, which is the case, e.g., for $\nu=0$.

\subsection{Criticality at tree-level}\label{ExplicitTree}
We begin by matching the full theory to the leading-order potential of the effective theory,
\begin{equation}\label{VIR}
    V_{\rm IR}(H)=\frac{1}{2} m^2_{\rm IR} H^2 +  \frac{1}{4} \lambda_{\rm IR} H^4 + \frac{C_6}{M^2}H^6+ ... \ .
\end{equation}
At tree-level, this can be achieved by inserting the solution for the equation of motion for $S$ into the full potential of Eq.~\eqref{VUV}. Expanding the result as a series in $H^2 / M^2$ and comparing with the potential of the effective theory in Eq.~\eqref{VIR},\footnote{Integrating out the heavy mode $S$ also induces terms $\propto H^2 \frac{\Box}{M^2}H^2$, which can be eliminated by a non-linear rescaling of $H$. We find that the resulting corrections to the coefficients vanish in near-critical scenarios. This can be traced back to the critical features' nature as local extrema, whose existence and type is invariant under the rescalings of interest here.} we find
\begin{gather}
    m^2_{\rm IR} = m_{H}^2, \ \ \ \lambda_{\rm IR}=\lambda_H - 2 \frac{\nu^2}{M^2} , \ \ \ C_6 = \frac{\kappa}{2}  \frac{\nu^2}{M^2}.
\end{gather}
These matching conditions next allow us to identify the critical point of the UV theory at tree-level,
\begin{gather}\label{treematch}
    m_{H}^2 \simeq 0, \ \ \ \lambda_H \simeq 2 \frac{\nu^2}{M^2} , \ \ \  \kappa  \frac{\nu^2}{M^2} \simeq \frac{M^2}{\mu_I^2} \frac{|\beta_\lambda |}{6 \sqrt{e}}.
\end{gather}
The first of these relations implies that $m_{H}^2\sim \text{(1-loop)}\cdot M^2$, which is necessary to cancel 1-loop corrections from integrating out the heavy mode. In contrast, the second relation allows for $\lambda_{\rm IR} \sim 0$ near the matching scale even for typical values of the tree-level coupling $\lambda_H$. 

The third relation, derived from the criticality of the dimension-six term, restricts the combination of UV parameters $\kappa \cdot \nu^2$ as a function of the matching scale as well as $\beta_\lambda$ and $\mu_I$, which are fully determined by the observed values of the related IR couplings. We note that the combination $\kappa \cdot \nu^2$ as a function of the matching scale $M$ is of the simple form 
\begin{equation}
    \kappa \cdot \frac{\nu^2}{M^2} = \frac{M^2}{M_0^2}, \ \ \ \text{where} \ \ \ M_0^2= \frac{6 \sqrt{e}}{|\beta_\lambda|}\cdot \mu_I^2.
\end{equation}
Repeating our numerical analysis of section~\ref{rangeofscales}, we find for the central value of this coefficient
\begin{equation}
    M_{0,\rm central}= 3 \cdot 10^{13}\, \text{GeV}.
\end{equation}
Meanwhile, the allowed ranges at 68\% and 95\% confidence are given by
\begin{gather} 
    5\cdot 10^{11}\, \text{GeV} < M_{0,68\%} < 3\cdot 10^{18}\, \text{GeV}, \nonumber \\ 
    M_{0,95\%} > 10^{11}\, \text{GeV} .
\end{gather}
We once again find a significant error bar, which arises predominantly from the experimental uncertainties associated with the IR matching conditions of Eq.~\eqref{treematch}. Adding right-handed neutrinos (RHNs) subject to a low-scale seesaw, the parameter $M_0$ can be lowered to energy scales that (in principle) could be accessible with realistic accelerators. Taking the most extreme case of RHN masses $M_\nu \sim \mathcal{O}(\text{TeV})$ and Yukawa couplings $|Y_\nu|\sim \mathcal{O}(1)$, their effect can lower the feasible values of this parameter down to values as low as $M_0 \sim 10^6 \, \text{GeV}$~\cite{Khoury:2021zao,Chauhan:2023pur}. Lastly, demanding perturbativity of the UV couplings provides an upper bound on the matching scale $M$: $M^2 < (4 \pi)^3 M_0^2$ for any given $M_0^2$.

The relation corresponding to a critical dimension-six term also determines the scale corresponding to the critical feature,
\begin{equation}\label{H2M2}
    \bar{H}^2  \simeq \frac{M^2}{\nu^2} \frac{|\beta_\lambda |}{6 \kappa} M^2.
\end{equation}
This relation can now be used to determine the limitations of our perturbative treatment, which can be infered from the condition that the dimension-six term be larger than higher-dimensional terms, and in particular the dimension-eight term. Having at hand a concrete theory it is straightforward to determine its respective Wilson coefficient,
\begin{equation}\label{C8}
    C_8= - \frac{\kappa^2}{2}   \frac{\nu^2}{M^2} + \frac{\lambda_S}{4} \frac{\nu^4}{M^4} .
\end{equation}
Using Eqs.~\eqref{H2M2} and~\eqref{C8}, the condition $|C_6| H^6/\Lambda^2 > |C_8| H^8/\Lambda^4$ can now be translated to a bound on the parameters of the UV theory in terms of the IR-beta function $\beta_\lambda$. Scanning over the entirety of the $95$\% error ellipsoid, we find that
\begin{equation}
    \kappa \cdot \frac{\nu^2}{M^2} \cdot \left| - \kappa + \frac{1}{2} \frac{\lambda_S}{\kappa} \frac{\nu^2}{M^2} \right|^{-1} > \frac{|\beta_\lambda|}{6} \gtrsim 3 \cdot 10^{-4},
\end{equation}
with even weaker bounds on the end of the error ellipsoid corresponding to larger instability scales. Even in the previously discussed scenario of right-handed neutrinos in a low-scale seesaw with $\mathcal{O}(1)$-Yukawa couplings, this bound remains of order $\mathcal{O}(10^{-2})$.

\begin{figure*}[t!]
    \centering
    \includegraphics[width=0.4\textwidth]{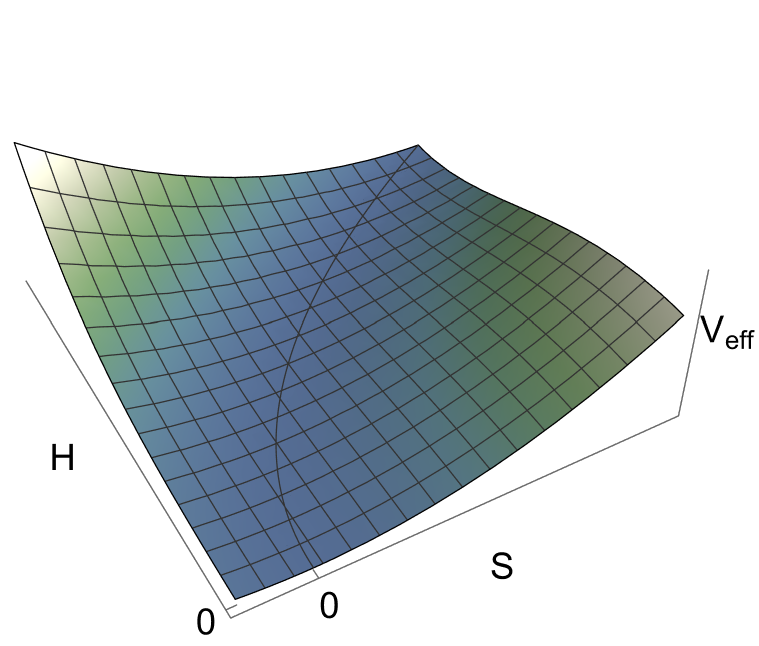}
    \includegraphics[width=0.4\textwidth]{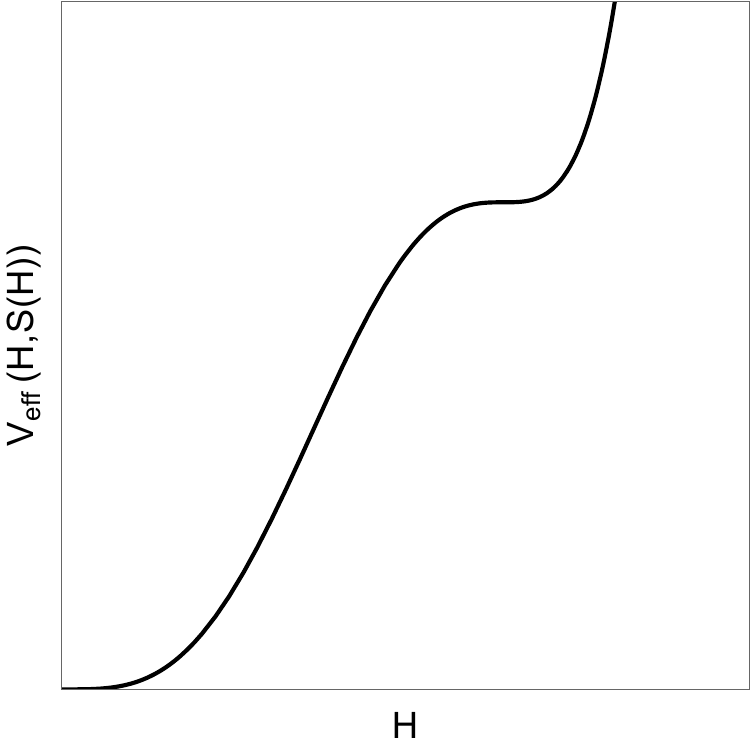}
 \caption{Left: The critical UV potential at tree-level for a generic set of UV parameters, taking into account only the running of $\lambda_H$. The black line represents the subspace on which the static equation of motion for $S$ is satisfied. The displacement along this line represents the effective field of the low energy theory. Right: The potential along the low-energy subspace highlighted by the black curve in the left panel. The critical feature is clearly visible.}\label{TreePlotFull}
\end{figure*}

The critical feature also has a clear interpretation from the perspective of the full theory. Integrating out the heavy mode $S$ amounts to restricting the dynamics to the relatively flat ``valley'' roughly parallel to the $H$-axis, as shown in Fig.~\ref{TreePlotFull}. For large values of $H$, the terms $\propto \nu , \kappa$ cause an excitation along the $S$-direction, which manifests as a positive contribution to the overall potential. If these terms are smaller than some critical value, this upward shift only becomes important after the running of $\lambda_H$ has driven the potential to negative values. For larger values of the $\nu$ and $\kappa$ terms, this effect becomes important for small enough values of $H$, preventing the formation of a critical feature. The critical behavior thus corresponds to the exact point at which the ``flipping'' of the $H^4$-term is precisely balanced by the positive contribution of the field $S$, which to leading order manifests through the dimension-six term. 

It is easy to see that such a behavior can be realized even if our perturbative treatment breaks down through the combined effect of the dimension-six as well as higher-dimensional terms, as long as the effect of the new physics does not prevent the quartic coupling from turning negative. The condition $\mu_I \lesssim M$ can thus be understood as a restriction of our perturbative treatment rather than a strict limit for the actual values of the physical scales. In contrast, the upper bound $M^2 < (4 \pi)^3 M_0^2$ has a clear physical interpretation, representing perturbativity of the UV theory.

\subsection{Criticality at one-loop}\label{ExplicitLoop}

In the case $\nu=0$, the leading-order contribution to the dimension-six term arises at the one-loop level. In general, the coefficients in the UV and IR theory are---in this scenario, at one-loop accuracy and near the instability scale where $\lambda\sim$ (1-loop)---related to one another through
\begin{gather}\label{matching}
    m_{\rm IR}^2=m_H^2- \frac{\kappa}{(4 \pi)^2} M^2, \ \ \ \lambda_{\rm IR}= \lambda_H , \nonumber \\ C_6 = \frac{\kappa^3}{12 (4 \pi)^2} ,
\end{gather}
where the threshold correction $\propto \kappa^2$ to the quartic coupling vanishes through cancellations.\footnote{Integrating out the heavy scalar field $S$ also induces a momentum-dependent term $\propto \frac{\kappa^2}{(4 \pi)^2} H^2 \frac{\Box}{M^2}H^2$, which, as in the previous case, vanishes to leading order for near-critical parameters.}

Criticality thus leaves us, in principle, with two free parameters, $|\beta_\lambda|$ and $M^2/ \bar{H}^2$, where $\beta_\lambda$ can again be considered a parameter determined by electroweak-scale measurements. Similarly, $\mu_I$ can for this purpose be considered as given, leaving only $M$ as a free parameter.

Using the matching relations of Eq.~\eqref{matching}, criticality of the UV potential corresponds to 
\begin{gather}
    m_H^2 (M)=\frac{\kappa (M)}{(4 \pi)^2} M^2, \ \ \ \ \ \lambda_{\rm IR}\sim \text{1-loop} , \\ \frac{\kappa^3(M)}{(4 \pi)^2} = \frac{M^2}{\mu_I^2} \frac{|\beta_\lambda |}{\sqrt{e}} = \frac{M^2}{\bar{H}^2} |\beta_\lambda |.
\end{gather}
While the relations for the quartic coupling and the mass parameter $m_H^2$ could have been deduced directly from the well-known values of all relevant couplings in the IR, demanding criticality of the dimension-six term fixes the coupling $\kappa$ of the $H^2 S^2$ term. We find a similarly simple structure as in subsection~\ref{ExplicitTree},
\begin{equation}
    \kappa(M) = \left( \frac{M^2}{M_1^2} \right)^{\frac{1}{3}}, \ \text{where} \ \ M_1^2 = \frac{\sqrt{e}}{(4 \pi)^2 |\beta_\lambda|} \cdot \mu_I^2.
\end{equation}
Once again repeating the numerical analysis of section~\ref{ExplicitTree}, we find for the central value of this parameter
\begin{equation}
    M_{1,\rm central} = 3\cdot 10^{10}\, \text{GeV},
\end{equation}
while the allowed ranges at different confidence levels are given by
\begin{gather}
    5\cdot 10^{8} \, \text{GeV} < M_{0,68\%} < 3\cdot 10^{15} \, \text{GeV}, \\ 
    M_{0,95\%} > 10^{8} \, \text{GeV} .
\end{gather}
This can, once again, be translated into an upper bound on the matching scale $M$ by demanding perturbativity of the UV coupling $\kappa$, which yields $M^2< (4 \pi)^3 M_1^2$.

Meanwhile, the applicability of the effective theory can again be used to deduce a lower bound on $\kappa$. The Wilson coefficient of the dimension-eight term is now given by
\begin{equation}
    C_8 = -\frac{\kappa^4}{ 48(4  \pi)^2}.
\end{equation}
Again demanding the dimension-six term to be larger than the dimension-eight term this then implies that
\begin{equation}
    \kappa > \frac{1}{2} \left[(4 \pi)^2 |\beta_\lambda| \right]^{\frac{1}{2}}.
\end{equation}
Repeating our numerical analysis, we find for the precise values of the bound
\begin{gather}
     \frac{1}{2}\left[(4 \pi^2) |\beta_\lambda| \right]^{\frac{1}{2}}_{\rm central}=0.25 ,\\ 
     \frac{1}{2}\left[(4 \pi^2) |\beta_\lambda| \right]^{\frac{1}{2}}_{68\%}\in [0.12,0.36], \\  \frac{1}{2} \left[(4 \pi^2) |\beta_\lambda| \right]^{\frac{1}{2}}_{95\%}<0.36. 
\end{gather}
We note that this bound is significantly stronger than its equivalent in subsection~\ref{ExplicitTree}, in agreement with our general discussion from section~\ref{CritPot}, where we concluded that large values of the Wilson coefficient $C_6$ provide the most reliable effective descriptions.

To understand this bound, we can consider the case in which it is violated, i.e., in which the critical feature is not reliably described by the effective theory. If indeed $\langle S \rangle =0$, the effect of the second scalar field on the Higgs field's potential is reduced to its loop contribution,
\begin{align}
    \Delta V_{\rm eff,S}=& \frac{1}{(4 \pi)^2} \left[ \ln \left( \frac{M^2 + \kappa H^2}{\mu^2} \right) - \frac{3}{2} \right] \\ 
    & \times \left( \frac{M^4}{4} + \frac{\kappa M^2}{2} H^2 + \frac{\lambda}{4} H^4 \right). 
    \label{dVSmain}
\end{align}
For $\kappa H^2\ll M^2$, where $S$ is integrated out, expanding this term leads to the higher-order operators and threshold corrections of Eq.~\eqref{VIR}. Meanwhile, for $\kappa H^2\gg M^2$ it can be partially absorbed into the RG running of the full theory's parameters. Most importantly, the resulting effective potential contains only terms up to fourth order in $H$, preventing the formation of the critical feature discussed throughout this article.\footnote{It is nevertheless possible for a critical point to form if the modified running is such that both $\lambda_H (\mu_I)\sim \beta_{\lambda_H}(\mu_I)\sim 0$ at one-loop order. This notion of criticality is usually invoked in the context of Higgs inflation~\cite{Bezrukov:2014bra,Hamada:2014wna,Iacobellis:2016eof,Enckell:2016xse,Rasanen:2017ivk,Enckell:2020lvn,Salvio:2018rv,Lee:2023wdm,Kawai:2023vjo}.}

To better understand this behavior, it is helpful to revisit the corresponding scenario $\bar{H}\sim M \sim \mu_I$ together with $\kappa \ll 1$. In general, the last relation justifies an expansion of the correction in Eq.~\eqref{dVSmain} as a series in $\kappa \bar{H}^2 / M^2$, even in the full theory for which $\bar{H}^2 \gtrsim M^2$. This is, however, not always possible, as the field value at the critical feature generally scales as $\bar{H}^2 \propto M^2 / C_6$, and $C_6$ is a function of the couplings. In the model at hand, we have $C_6 \propto \kappa^3$, which overcompensates the small factor of $\kappa$ within the expansion parameter.

\section{Achieving (near-)criticality}\label{SOL&MV}

While the emergence of critical parameters is at least puzzling in the context of conventional particle physics model-building, it is a common feature of dynamical systems. In the context of the Higgs sector, this can been understood as a hint that its parameters are not fundamental constants, but rather arise through the coupling to some dynamical sector, e.g. involving one or more additional scalar fields. In this section, we review two qualitatively different examples for such mechanisms, unfolding during inflation and on the larger timescales of the multiverse. 

\subsection{Self-organized localization}

The central idea of \textit{self-organized localization} (SOL) is that critical points in the effective potential $V_{\rm eff} (H)$ serve as global attractors during early-universe inflation. A concrete realization of this idea has been proposed in Ref.~\cite{Giudice:2021viw}, which we summarize in the context of our conjecture.

Just as in our work, SOL sets out from the hypothesis that the near-critical parameters of the Higgs potential are functions of some additional light scalar field(s) undergoing large fluctuations during inflation. It is furthermore assumed that these fields are subject to some non-trivial vacuum structure, such that the number and properties of these fields' vacua depend on the Higgs field's parameters. It is then feasible that the dynamics of the additional fields are such that their probability distributions peak around values of the Higgs field's parameters corresponding to the coexistence of two distinct phases.

Taking into account only terms up to dimension four in the effective potential, the SOL hypothesis can account for the metastability of the electroweak vacuum, as the latter is necessary for the existence of minima in addition to the electroweak vacuum. Crucially for our result, however, the extrapolated RG-trajectory of the quartic coupling does \textit{not} imply the existence of such a vacuum \textit{at sub-Planckian scales}, as the minimum of the effective potential lies beyond $M_{\text{pl}}$ for all values of the couplings near their central values. Within the context of SOL, this could easily be fixed through the effects of additional bosons on the running of $\lambda$, but introducing such new fields would inevitably also induce a dimension-six term in the effective potential for the Higgs field. Moreover, if one insists on requiring that two distinct phases coexist within the effective potential (central to SOL), then the coefficient of the induced dimension-six term should assume a near-critical value. 

The dimension-six term is also required by the mechanism potentially responsible for determining the Higgs field's mass. As pointed out in Ref.~\cite{Giudice:2021viw}, the requirement of two coexisting phases suggests that $m_{\rm eff}$ should assume a value close to the metastability bound given in Eq.~\eqref{msbound}. As pointed out in section~\ref{rangeofscales}, the only possibility for this to occur, given the measured values of the couplings, is through the effect of some additional, as-yet unobserved fermions, such as right-handed neutrinos. Such new fermionic couplings, however, would destablize the vacuum, implying a vacuum lifetime significantly shorter than the current age of the universe. Avoiding this tension, in turn, requires the introduction of additional bosonic physics, again implying the existence of a dimension-six term.

\subsection{Multiverse selection}

A related but independent approach to vacuum selection relies on statistics of the vacuum landscape itself, which is assumed to be populated through eternal inflation~\cite{Khoury:2019yoo,Khoury:2019ajl,Kartvelishvili:2020thd,Khoury:2021zao}. A commonly considered scenario is that our universe exists an exponentially long time after the onset of eternal inflation, which would imply that the notion of a likely vacuum would have to be defined based on some near-equilibrium probability distribution, although, to date, all such discussions are limited by the \textit{measure problem}. An alternative scenario considered in Ref.~\cite{Khoury:2019yoo} is that we instead live in the early times of eternal inflation, in which it is possible to define the \textit{accessibility measure}, which quantifies how quickly a vacuum can be accessed~\cite{Khoury:2019ajl}. 

A central result of this approach is that it favors relatively shortlived vacua, with lifetimes of order of the \textit{Page time}, $\tau_{\text{Page}}\sim M_{\rm pl}^2 / H_0^3 \sim 10^{130}$~years (where $H_0$ denotes the present value of the Hubble parameter), compared to the SM central value~$\tau\sim 10^{983}$~years~\cite{Khoury:2021zao,Steingasser:2022yqx}. The requirement of a metastable vacuum furthermore implies a small Higgs mass through the metastability bound of Eq.~\eqref{msbound}. Just as in the case of SOL, these mismatches can be understood to suggest the existence of additional fermionic degrees of freedom, capable of lowering the instability scale as well as the lifetime of the vacuum. As noted above, however, lowering the metastability bound to values near the electroweak scale would destabilize the vacuum too much, which in turn would require additional bosonic physics capable of partially stabilizing the vacuum; such new degrees of freedom would typically manifest through a dimension-six operator in the effective potential. Thus, the value of the coefficient $C_6 / \Lambda^2$ of the induced dimension-six term would need to be large enough to compensate for a large destabilizing effect of the additional fermions, but 
not so large as to fully stabilize the vacuum. This, once again, can be understood as requiring a near-critical value for $C_6 / \Lambda^2$.

An alternative approach based on Bayesian reasoning has been put forward in Ref.~\cite{Khoury:2022ish}. That model features a series of local minima nearly degenerate with the electroweak vacuum. Such features arise as a special case of the near-critical behavior we discussed in section~\ref{NearCritPot}.

\section{Conclusion}

Both parameters in the Higgs field's potential, its mass and quartic self-coupling, appear fine-tuned to values near critical points that mark the transition between qualitatively different types of effective potentials. Such behavior is a common feature of dynamical systems, which has motivated many attempts to explain such peculiar behavior based on dynamical-selection mechanisms. Taking seriously the idea that such a mechanism might be realized in nature poses the question of how it would affect as-yet unobserved BSM physics. 

An obvious possibility is that the dynamical selection of critical values for various parameters also applies to new physics, which allows concrete predictions to be made regarding novel phenomena. Using an effective field theory description, we have identified such a critical point for a wide class of SM extensions. This critical point corresponds to a tuning of the coefficient of the dimension-six term in conjunction with the RG trajectory of the quartic coupling $\lambda$ and manifests through the formation of a saddle point in the effective potential at a particular scale. 

Remarkably, such tuning yields novel features in the effective potential at energy scales well below the scale $\Lambda$ at which one might expect BSM contributions to become important. In general, we find that the critical feature manifests near the instability scale $\mu_I$, at which the one-loop-corrected SM potential changes sign, defined via $\lambda_{\rm eff} (\mu_I, H = \mu_I) = 0$; we quantify both the central value for $\mu_I$ and its allowable range, based on present measurements of relevant SM couplings and masses. We find that the requirement of perturbativity in the UV theory translates to an upper bound on the scale of new physics, while the very existence of the dimension-six term can, in many cases, be translated to a lower bound. Plausible scenarios---such as a combination of heavy right-handed neutrinos plus a near-critical dimension-six term in the Higgs potential---suggest that the scale of new physics could be as low as $\Lambda \sim {\cal O} (10^6 - 10^9) \, {\rm GeV}$. On the other hand, a region of allowed parameter space corresponds to $\Lambda \sim M_{\rm pl}$, which could have implications for cosmological scenarios such as Higgs inflation. For the central values of the relevant parameters, the relevant feature lies within the range of typical post-inflation reheating temperatures, allowing for the possibility of a first-order phase transition during the early universe. We illustrate this range with the concrete example of a singlet extension of the SM, both with and without a tree-level mixing of the Higgs field with the new, heavy scalar field.

The notion of the dimension-six term being near-critical also offers a new perspective on previous works. First, it allows for a natural realization of the Multiple Point Principle~\cite{Froggatt:1995rt,Froggatt:2001pa,Das:2005ku,Froggatt:2005nb,Das:2005eb,Froggatt:2004nn,Froggatt:2006zc,Froggatt:2008hc,Das:2008an,Klinkhamer:2013sos,Froggatt:2014jza,Haba:2014sia,Haba:2014qca,Kawana:2014zxa,Hamada:2015fma,Hamada:2015ria,Haba:2016gqx,Haba:2017quk,McDowall:2018tdg,McDowall:2019knq,Maniatis:2020wfz,Kannike:2020qtw,Kawai:2021lam,Hamada:2014xka,Haba:2014zda,Kawana:2015tka,Hamada:2021jls,Racioppi:2021ynx,Huitu:2022fcw,Hamada:2015dja,Hamada:2017yji,Kawana:2016tkw}, which assumes that some special scale $\mu$ exists at which $V_{\rm eff}(\mu)=V_{\rm eff}(v_{\rm EW})$; the absolute stability scale for the Higgs potential can be considered as a special case of such a scale. We further demonstrate that a tuned, near-critical dimension-six term in the effective potential is a natural consequence of several vacuum selection models, which favor a lowering of the instability scale compared to the SM value~\cite{Giudice:2021viw,Khoury:2019yoo,Khoury:2019ajl,Kartvelishvili:2020thd,Khoury:2021zao,Khoury:2022ish}. 

Another interesting possibility would be to explicitly investigate the interplay between the dimension-six term with other mechanisms that favor near-criticality. Particularly relevant are models that invoke \textit{cosmological relaxation}~\cite{Arkani-Hamed:2020yna,Csaki:2020zqz,Strumia:2020bdy,Cheung:2018xnu,Graham:2015cka,Nelson:2017cfv,Gupta:2018wif,TitoDAgnolo:2021pjo,TitoDAgnolo:2021nhd,Trifinopoulos:2022tfx}. These are usually based on some dynamical back-reaction of the Higgs-vev on some additional scalar field, whose properties are such that regions of space in which the Higgs mass takes a large value undergo rapid collapse. Allowing for a second minimum in the Higgs field's effective potential thus allows for a multi-staged selection process, so long as the Higgs field is temporarily trapped in the second minimum~\cite{Rafelski:2015lva}.

The ongoing tensions between observations and the well-established principles of naturalness and symmetry pose a significant challenge for particle physics model-building beyond the Standard Model. Whereas many attempts have been made to bring forward new guiding principles, most specific predictions rely on concrete models, which are often based on non-trivial assumptions. In comparison, our results have broad applicability, and can be understood as a first step towards a more unified perspective on dynamical-selection models. By exploiting EFT techniques to explore consequences of a near-critical effective potential, our approach builds on a minimal set of assumptions and remains independent of concrete mechanisms that might underlie any such dynamical-selection processes. As such, it may help guide the search for as-yet unobserved BSM phenomena.

\section*{Acknowledgements}
We gratefully acknowledge helpful discussions with Jos\'e R. Espinosa, Sarah Geller, Matthew McCullough, Jesse Thaler, and Jan-Niklas Toelstede.
TS's contributions to this work were made possible by the Walter Benjamin Programme of the Deutsche Forschungsgemeinschaft (DFG, German Research Foundation) -- 512630918. Portions of this work were conducted in MIT's Center for Theoretical Physics and partially supported by the U.S. Department of Energy under Contract No.~DE-SC0012567. This project was also supported in part by the Black Hole Initiative at Harvard University, with support from the Gordon and Betty Moore Foundation and the John Templeton Foundation. The opinions expressed in this publication are those of the author(s) and do not necessarily reflect the views of these Foundations.

\bibliography{ref}

\end{document}